\documentclass[preprintnumbers,nofootinbib,a4paper,10pt]{revtex4-1}

\usepackage[utf8x]{inputenc}

\usepackage{changepage}
\usepackage{setspace} %double space?
\usepackage{microtype} %textls[-15] ?

\usepackage{amsmath,latexsym,amssymb,amsfonts}
\usepackage{float} \usepackage{graphicx}% Include figure files
\usepackage{epstopdf}
\usepackage{graphics}
\usepackage{caption}
\usepackage{subfigure}
\usepackage{bm}
\usepackage{color}
\usepackage{hyperref}
\hypersetup{
	colorlinks=true,
	linkcolor=blue,
	filecolor=magneta,      
	urlcolor=blue,
}
\usepackage[dvips]{psfrag}
\usepackage{ulem}
\usepackage{epsfig}
\usepackage{cases}
\usepackage{tensor}
\usepackage{multirow,array}
\usepackage{booktabs} %For prettier tables

\usepackage[dvipsnames]{xcolor}

\usepackage{colortbl} %collored table

\def\lambdabar{\ThisStyle{\ensurestackMath{\stackon[-2.4\LMpt]{%
  \SavedStyle\lambda}{\kern-.5pt\kern\LMpt\rule{1\LMex}{.25pt+.15\LMpt}}}}}

\newcommand{\m}{\text{m}}

\newcommand{\tc}{\tilde{c}}

\newcommand{\TM}{\text{M}}
\newcommand{\TL}{\text{L}}

\newcommand{\Tth}{\text{th}}

\newcommand{\Tm}{\text{m}}

\newcommand{\SMC}{\text{SMC}}

\newcommand{\TMpc}{\text{Mpc}}

\newcommand{\Tobs}{\text{obs}}

\begin{document}

\title{Constraints on the minimally extended varying speed of light model using Pantheon+ dataset} %%%%%%%%%%%%
\author{Seokcheon Lee}
%\date{\today}
\email[Email:]{skylee2@gmail.com}
\affiliation{Department of Physics, Institute of Basic Science, Sungkyunkwan University, Suwon 16419, Korea}

%\let\thefootnote\relax
%\footnotetext{MSC2020: Primary 00A05, Secondary 00A66.} %%%%%%%%%%

\begin{abstract}
In the context of the minimally extended varying speed of light (meVSL) model, both the absolute magnitude and the luminosity distance of type Ia supernovae (SNe Ia) deviate from those predicted by general relativity (GR). Using data from the Pantheon$+$ survey, we assess the plausibility of various dark energy models within the framework of meVSL. Both the constant equation of state (EoS) of the dark energy model ($\omega$CDM) and the Chevallier-Polarski-Linder (CPL) parameterization model ($\omega = \omega_0 + \omega_a (1-a)$) indicate potential variations in the cosmic speed of light at the 1-$\sigma$ confidence level. For $\Omega_{\m0} = 0.30, 0.31$, and 0.32 with $(\omega_0 \,, \omega_a) = (-1 \,, 0)$, the 1-$\sigma$ range of $\dot{c}_0/c_0 \, (10^{-13} \, \text{yr}^{-1}) $ is (-8.76 \,, -0.89), (-11.8 \,, 3.93), and (-14.8 \,, -6.98), respectively. Meanwhile, the 1-$\sigma$ range of $\dot{c}_0/c_0 (10^{-12} \, \text{yr}^{-1}) $ for CPL dark energy models with $-1.05 \leq \omega_{0} \leq -0.95$ and $0.28 \leq \Omega_{\m0} \leq 0.32$, is (-6.31\,, -2.98). The value of $c$ at $z = 3$ can exceed that of the present by $0.2 \sim 3$ \% for $\omega$CDM models and $5 \sim 13$ \% for CPL models. Additionally, for viable models except for the CPL model with $\Omega_{\m0} = 0.28$, we find $-25.6 \leq \dot{G}_0/G_0 \, (10^{-12} \, \text{yr}^{-1}) \leq -0.36$. For this particular model, we obtain an increasing rate of the gravitational constant within the range $1.65 \leq \dot{G}_0/G_0 \, (10^{-12} \, \text{yr}^{-1}) \leq 3.79$. We obtained some models that do not require dark matter energy density through statistical interpretation. However, this is merely an effect of the degeneracy between model parameters and energy density and does not imply that dark matter is unnecessary.
\end{abstract} %%%%%%%%%

\maketitle

\section{Introduction}
\label{sec:intro}

The principles of physics should remain unchanged regardless of the chosen units or measurement instruments. This is exemplified by dimensionless quantities such as the fine structure constant, $\alpha$, which retains its value across different unit systems, as demonstrated in the Standard Model of particle physics. Conversely, dimensional constants like $\hbar$, $c$, $G$, $e$, and $k$ are human conventions whose numerical values vary depending on the units employed. Consequently, only dimensionless constants can be considered truly fundamental. Therefore, investigating potential variations over time in dimensionless fundamental constants is a valid scientific pursuit, while variations in dimensional constants like $c$ or $G$ depend on the chosen units and may result in discrepancies among observers. This perspective is supported by various studies~\cite{Duff:2001ba,Uzan:2002vq,Ellis:2003pw,Duff:2014mva}, albeit within the framework of a static Universe or one existing at the present epoch \cite{Lee:2020zts,Lee:2023bjz,Lee:2024par}.

In the Robertson-Walker (RW) metric, the expanding Universe is depicted as progressing from one hypersurface to another, with the scale factor increasing naturally, resulting in the cosmological redshift of various physical quantities, such as mass density, wavelength, and temperature. One can derive this metric from the cosmological principle and Weyl's postulate. The redshift, defined as a function of the time-evolving cosmic scale factor $a(t)$, yields positive values for $z$ in our expanding Universe. Estimating the redshift of a galaxy involves analyzing the emission lines emitted by glowing gas within the galaxy. For example, the H$\alpha$ line, a red Balmer line of neutral hydrogen, has a rest wavelength of $6562$\AA. If the observed wavelength of this line presently measures $8100$\AA, it indicates that the galaxy is positioned at $z = 0.234$ (i.e., $a = 0.81$). Consequently, in an expanding Universe, the value of dimensional quantities, such as wavelength, varies depending on the observation time (i.e., cosmic time). Additionally, it has been observed that the temperature of the CMB decreases with the age of the Universe, scaling inversely with the scale factor as $T = T_0 a^{-1}$ \cite{Islam01,Narlikar02,Hobson06,Roos15}. However, the RW model lacks a mechanism to determine cosmological time dilation (TD). The standard model of cosmology (SMC) makes an additional assumption, asserting that the speed of light is constant ($c$). This assumption arises from the dependence of SMC on general relativity (GR), which assumes that $c$ is invariant. As a result, the cosmological TD between two hypersurfaces at $t = t_1$ and $t = t_2$ vary in proportion to the inverse of the scale factors $a(t)$ at those specific times. However, if one allows a time-varying speed of light as proposed in this paper, this relationship may not hold anymore. Thus, establishing TD depends on experimental observations. Given the theoretical absence of cosmological TD, the relationship can be considered as a function of the scale factor, allowing the speed of light to be expressed as $c(t_1)=  (a(t_1)/a(t_2))^{b/4} c(t_2)$ in the meVSL model, where $b$ is a constant \cite{Lee:2020zts,Lee:2023bjz,Lee:2024par}. Also, 
to preserve the Einstein Field Equations (EFE), the Einstein constant, denoted as $\kappa = 8 \pi G/c^4$, must remain constant even if both $c$ and $G$ undergo cosmological evolution. In the meVSL model, where $c = c_0 a^{b/4}$, this implies $G = G_0 a^{b}$ where $c_0$ and $G_0$ represent the current values of the speed of light and the gravitational constant, respectively \cite{Lee:2020zts,Lee:2023bjz,Lee:2024par}.

In addition to models like meVSL, there is a model such as Co-varying Physical Couplings (CPC)  \cite{Cuzinatto:2022mfe,Cuzinatto:2022vvy,Cuzinatto:2022dta}, where physical constants vary with cosmic time. The CPC retains the EFEs with $G$, $c$, and $\Lambda$ treated as functions of spacetime. The interaction between the Bianchi identity and the requirement of stress-energy tensor conservation complicates the potential variations of these constants, which are constrained to co-vary according to the General Constraint (GC). Unlike meVSL, this model includes the dynamics of physical constants through the adoption of GC. 

Special relativity (SR)'s universal Lorentz covariance, grounded in Minkowski spacetime, adequately upholds its principles \cite{Morin07}. In contrast, within GR, an inertial frame (IF) refers to one that is freely falling. While Lorentz invariant (LI) spacetime intervals can be established between events, the definition of a global time in GR is impeded by the absence of a universal IF. However, a global time can be delineated for the Universe, satisfying the Cosmological Principle (CP), enabling a foliation of spacetime into non-intersecting spacelike 3D surfaces. This description pertains to the Universe modeled by the RW metric \cite{Islam01,Narlikar02,Hobson06,Roos15}. The LI varying speed of light (VSL) model is plausible if the speed of light, denoted as $c$, remains locally constant at each given epoch but varies in cosmic time \cite{Lee:2020zts,Lee:2023bjz,Lee:2024par}. In other words, in an expanding Universe, if the speed of light is expressed as a function of the scale factor, $c[a]$, then although its value changes akin to wavelengths at different epochs, such as $a_1$ and $a_2$, it maintains a constant local value at each epoch, ensuring LI and thereby preserving the validity of quantum mechanics and electromagnetism in accordance with SR during every epoch. Thus, the speed of light could change over cosmic time or remain constant within an expanding Universe, contingent upon its relationship with cosmological TD. Without explicit laws governing TD, the speed of light in the RW metric could potentially vary with cosmological time, akin to other physical properties such as mass density, temperature, and fundamental constants like the Planck constant \cite{Lee:2020zts,Lee:2022heb,Lee:2023bjz,Lee:2024par}. However, to construct a coherent model around this concept, the varying speed of light (VSL) needs to be incorporated into the Einstein field equations (EFEs) and resolved for solutions. Previous studies, notably within the framework of the minimally extended VSL (meVSL) model, have addressed such scenarios \cite{Lee:2020zts,Lee:2023bjz,Lee:2024par}. One can freely select a local value for the speed of light as it merely entails a scaling of length units. As long as this local value remains constant on a given time hypersurface, it satisfies the SR to be consistent with local physics laws. Newton's gravitational constant, $G$, could potentially vary. To avoid trivial unit rescaling, one must examine the concurrent variation of $c$, $G$, and possibly other physical constants \cite{Lee:2020zts,Lee:2023bjz,Lee:2024par}. 

Numerous endeavors have aimed to measure cosmological TD. One approach involves directly observing TD by analyzing the decay time of distant supernova (SN) light curves and spectra~\cite{Leibundgut:1996qm,SupernovaSearchTeam:1997gem,Foley:2005qu,Blondin:2007ua,Blondin:2008mz}. Another method entails measuring TD by examining the stretching of peak-to-peak timescales of gamma-ray bursters (GRBs) \cite{Norris:1993hda,Wijers:1994qf,Band:1994ee,Meszaros:1995gj,Lee:1996zu,Chang:2001fy,Crawford:2009be,Zhang:2013yna,Singh:2021jgr}. Efforts have also been made to detect the TD effect in the light curves of quasars (QSOs) located at cosmological distances~\cite{Hawkins:2001be,Dai:2012wp}. So far, no definitive detection of cosmic TD has been achieved, with conflicting results from different measurements. 

The assessment of whether the speed of light alters due to the Universe's expansion hinges solely on observation. Therefore, it is vital to observe how any changes in the speed of light affect cosmic scales. To achieve this, diverse observational methods have been utilized  \cite{Lee:2020zts,Lee:2024par}.  

First, there is the cosmic distance duality relation CDDR  method among these methods. Etherington's theorem, derived from the geodesic deviation equation, establishes reciprocity between the area distances of galaxies and observers, linked by the redshift factor $(1+z)$ under geometric invariance~\cite{Etherington:1933pm}. This theorem, applicable in spacetimes where photons follow null geodesics, forms the basis for the CDDR. By relating area distances to angular and luminosity distances, the CDDR offers a means to test the validity of the SMC ~\cite{Ellis:1998ct,Ellis:2007grg}. Various tests of the CDDR using astrophysical and cosmological observations have been conducted to constrain VSL models \cite{More:2008uq,Nair:2012dc,Wu:2015prd,Ma:2016bjt,Martinelli:2020hud,Holanda:2012ia,Qi:2014zja,Salzano:2014lra,Lee:2021xwh,Rodrigues:2021wyk,Cuzinatto:2022mfe}. Our analysis of the meVSL model suggests a potential deviation from the standard CDDR based on current data. However, with different priors for certain cosmological parameters, the current dataset aligns with the SMC, indicating no deviation from the expected CDDR~\cite{Lee:2020zts,Lee:2021xwh}. Therefore, acquiring more precise data is essential to thoroughly investigate any deviations from the established CDDR and reaffirm the viability of the meVSL model. 

Second, there is the Cosmic Chronometer (CC) method. The CC method involves observing two passively evolving galaxies, typically elliptical galaxies, assumed to have formed at the same cosmic epoch but observed at different redshifts ~\cite{Jimenez:2001gg}. This approach offers a model-independent means of measuring the Hubble parameter, $H(z)$, as a function of redshift, derived from spectroscopic surveys with high precision ($\sigma_z \leq 0.001$). The expansion rate of the meVSL model, or the Hubble parameter $H(z)$, is determined from the differential age evolution of the Universe $\Delta t$ within a given redshift interval ($dz$) \cite{Lee:2020zts,Lee:2023bjz,Lee:2024par}
\begin{align}
H(z) \equiv \frac{\dot{a}}{a} &= - \frac{1}{1+z} \frac{dz}{dt} \approx - \frac{1}{1+z} \frac{\Delta z}{\Delta t} = H(z)^{(\textrm{SMC})} (1+z)^{-b/4} \nonumber \\ &= H_0 E(z)^{(\textrm{SMC})} (1+z)^{-b/4} \label{HzCC} \,,
\end{align}
where $E^{(\textrm{SMC})}$ is the normalized Hubble parameter in the SMC model. 
Various methods exist for measuring $\Delta t$, including predicting its age based on the chemical composition of a stellar population or utilizing spectroscopic observables like the $4000$ \AA break, known to be linearly related to the age of the stellar population~\cite{Moresco:2010wh}. Unlike many cosmological measurements that rely on integrated distances, the CC method determines the expansion rate $H(z)$ as a function of the redshift–time derivative $dz/dt$, making it a potent tool for testing different cosmological models~\cite{Wei:2016ygr,Ratsimbazafy:2017vga,Wei:2019uss,Moresco:2020fbm,Vagnozzi:2020dfn,Dhawan:2021mel,Borghi:2021zsr,Borghi:2021rft,Banerjee:2022ynv,Jalilvand:2022lfb,Asimakis:2022jel,Kumar:2022ypo,Li:2022cbk}.. This method proves particularly valuable for investigating VSL models~\cite{Rodrigues:2021wyk}. Both minimum $\chi^2$-analysis and maximum-likelihood analysis using the most recent CC data have been performed to constrain the parameter b of the meVSL model. However, the precision of the current CC data is insufficient to distinguish between the meVSL model and the SMC~\cite{Lee:2023rqv}.

Third, there are cosmological TDs observed from supernovae. The luminosity curve (LC) of a supernova (SN) offers valuable insights into its evolution, aiding in the classification and understanding of its properties. LC analysis helps determine crucial parameters such as peak luminosity, time to peak brightness, and rate of decline, particularly for SNe Ia, essential as standard candles in cosmology. Comparing LCs across distances enables investigation into cosmic expansion and TD, contributing to significant discoveries like accelerated expansion and dark energy. Wilson's method involves comparing LCs of nearby and distant SNe, revealing TD effects due to light travel time through space~\cite{Wilson:39}. This information is crucial for studying the Universe's expansion rate and testing cosmological models, involving data collection, mathematical modeling, and comparison of observed TD with theoretical predictions. We derived a TD formula within the meVSL model, analyzing data from $13$ high-redshift SNe Ia to determine the exponent $b$ as $b = 0.198 \pm 0.415$~\cite{Blondin:2008mz,Lee:2023ucu}. While less precise than CC, our analysis indicates consistency with both SMC and the meVSL model. Thus, distinguishing between the two based on SNe TD data is challenging.

Fourth, there is the cosmography method. It employs a kinematic description of the Universe's evolution based on the cosmological principle, emphasizing the dynamics of cosmic expansion. As a model-independent framework, it offers flexibility in managing cosmological parameters, allowing for generalized analysis unconstrained by preconceived models. By focusing on the later stages of cosmic evolution and utilizing Taylor expansions tailored to the observable domain where $z \ll 1$, cosmography imposes constraints on the present-day Universe. We adapt late-time cosmography to incorporate meVSL models~\cite{SLee:24CG}.

Lastly, in this paper, we aim to discuss the constraint on any evidence of the cosmic variation of the speed of light using Pantheon$+$ data~\cite{Scolnic:2017caz}. The Pantheon compilation comprises a total of $1048$ SNe Ia spanning a redshift range of $0.01$ to $2.3$. It includes $365$ spectroscopically confirmed SNe Ia from the Pan-STARRS1 (PS1) Medium Deep Survey, combined with a subset of $279$ PS1 SNe Ia (with redshifts ranging from $0.03$ to $0.68$) with reliable distance estimates obtained from various sources such as SDSS, SNLS, and HST samples. Cosmological models fitted to minimize the $\chi^2$ for flat $\Lambda$CDM and $\omega$CDM models, without accounting for systematic uncertainties on $\Omega_{m0}$, yield values of $0.284 \pm 0.012$ and $0.350 \pm 0.035$, respectively. The $1$-$\sigma$ constraint on $\omega$ for the $\omega$CDM model is $-1.251 \pm 0.144$. The Pantheon dataset allows for a precise constraint of approximately 10\% on $\Omega_{m0}$ and 12\% on $\omega$ for the flat $\omega$CDM model, and about 4\% on $\Omega_{m0}$ for the flat $\Lambda$CDM model.

In Section~\ref{sec:meVSL}, we give a brief overview of the meVSL model, delving into its theoretical foundations and implications for cosmological phenomena. Specifically, we explore the luminosity distance predictions derived from this model, shedding light on its unique characteristics and potential effects for observational data. In Section \ref{sec:chi2}, we elucidate the statistical methods employed in this paper. Additionally, a portion of the Mathematica file used for this purpose is included in the appendix. In Section \ref{sec:dotc}, we provide a brief explanation of the temporal variations of the speed of light and Newton's constant used in this paper, along with a concise interpretation for the $\Lambda$CDM model for comparison. The investigation into models capable of implementing variations in the speed of light across various $\omega$CDM models is conducted in Section \ref{sec:wCDM}. Research on potential models within the CPL framework of the meVSL model is performed in Section \ref{sec:CPL}. In Section \ref{sec:dotcdotG}, we offer various possible constraints on the temporal variations of the speed of light and Newton's constant within the meVSL model, comparing them with existing observational constraints. Finally, in Section~\ref{sec:conclusion}, we distill our findings and insights into a comprehensive summary, drawing actionable conclusions and outlining avenues for future research and exploration in the realm of cosmology and fundamental physics.

%%%%%%%%%%%%%%%%%%%%%%%%%%%%%%%%%%%%%%%%%%
\section{Summary for the meVSL}
\label{sec:meVSL}

The conceptualization of the four-dimensional spacetime of a spatially homogeneous and isotropic, expanding universe entails envisioning it as a seamless continuum composed of homogeneous and isotropic spatial hypersurfaces evolving dynamically over cosmic time \cite{Islam01,Narlikar02,Hobson06,Roos15}. At the heart of this framework lies the RW metric, elegantly expressed by 
\begin{align}
ds^2 &= - c(t)^2 dt^2 + a^{2}(t) \gamma_{ij} dx^i dx^j = -c(t)^2 dt^2 + a^2(t) \left( \frac{dr^2}{1-kr^2}  + r^2 d \Omega^2 \right) \nonumber \\ &= - c(t)^2 dt^2 + a^2(t) \left[ d \chi^2 + f_{k}^2(\chi) d \Omega^2 \right] \equiv - c(t)^2 dt^2 + a(t)^2 dl_{3\textrm{D}}^2  \label{ds2} \,,
\end{align}
where $c(t)$ denotes the speed of light and $\gamma_{ij}(\vec{x})$ signifies the time-independent spatial metric defining the hypersurface, while $a(t)$ governs the scale factor dictating the relationship between physical distance and comoving distance. We adopt that the speed of light is a function of cosmic time, deviating from the conventional RW metric. We already show that there is no contradiction in this assumption as long as we adopt the CP and Weyl's postulate \cite{Lee:2023bjz,Lee:2024par}.  The derivation of redshift involves utilizing the geodesic equation for a light wave, where $ds^2 = 0$ as represented by Equation~\eqref{ds2}. The consistency of $dl_{3\textrm{D}}$ over time is maintained by exclusively employing comoving coordinates. Building upon this foundation, we arrive at the expression for radial light signals:
\begin{align}
d l_{3\textrm{D}} &= \frac{c(t_i) dt_i}{a(t_i)} \quad : \quad \frac{c_1 dt_1}{a_1} = \frac{c_2 dt_2}{a_2} \Rightarrow \begin{cases} c_1 = c_2 = c & \textrm{if} \quad \frac{dt_1}{a_1} = \frac{dt_2}{a_2} \quad \textrm{SMC}  \\ c_1 = \left( \frac{a_1}{a_2}\right)^{\frac{b}{4}} c_2 & \textrm{if} \quad \frac{dt_1}{a_1^{1-\frac{b}{4}}} = \frac{dt_2}{a_2^{1-\frac{b}{4}}} \quad \textrm{meVSL}  \end{cases} \,, \label{dl3D}
\end{align}
where $d t_i = 1/\nu(t_i)$ denotes the time interval between successive crests of light at $t_i$ (\textit{i.e.}, the inverse of the frequency $\nu_i$ at $t_i$) and $b$ characterizes the deviation of $c$ from the constant value. Thus, the RW metric naturally allows the VSL models if we remove the traditional assumption on the cosmological TD \cite{Lee:2020zts,Lee:2023bjz,Lee:2024par}. 

Additionally, the introduction of $\chi = D_{\text{C}}$ as the comoving distance as  
 \begin{align}
 D_{\text{C}}(z) \equiv \int_{0}^{r} \frac{dr'}{\sqrt{1-kr^{'2}}} = \frac{c_0}{H_0} \int_{0}^{z} \frac{dz'}{E^{(\textrm{SMC})}(z')} \equiv \frac{c_0}{H_0} d_{\text{C}}(z) %=  D_{\text{C}}^{(\GR)}(z)
 \label{Dcmp} \,,
 \end{align}
where $d_{\TL}$ is the so-called the Hubble free luminosity distance and $f_{k}(\chi) = \sinh (\sqrt{-k} \chi )/\sqrt{-k} = D_{\TM}$ as the transverse comoving distance 
\begin{align}
 D_{\text{M}}(z) = D_{\text{M}}^{(\textrm{SMC})}(z) = \begin{cases} \frac{c_0}{H_0} \frac{1}{\sqrt{\Omega_{k0}}} \sinh \left( \sqrt{\Omega_{k0}} \frac{H_0}{\tc_0} D_{\text{C}} \right) & \Omega_{k0} > 0 \, \\D_{\text{C}} & \Omega_{k0} = 0 \, \\ \frac{c_0}{H_0} \frac{1}{\sqrt{|\Omega_{k0}|}} \sin \left( \sqrt{|\Omega_{k0}|} \frac{H_0}{\tc_0} D_{\text{C}} \right) & \Omega_{k0} < 0 \, \end{cases} \label{DMmp} \,,
 \end{align}
with $c_0 = 2.9979 \times 10^{5}$ km/s and $H_0 = 100 \textrm{h}$ km/Mpc/s representing the present values of the speed of light and the Hubble parameter, respectively~\cite{SLee:24CG,Tiesinga:2022}. In a recent local measurement by the SH0ES collaboration, utilizing Cepheid-calibrated SNeIa, the Hubble constant was reported to be approximately $73$ km/s/Mpc \cite{Breuval:2024lsv}, contrasting with the prediction of around $67$ km/s/Mpc by the standard $\Lambda$CDM model based on observations of the CMB \cite{Tristram:2023haj}.

Within this conceptual framework, the timelike worldlines of constant space delineate the threading, while the spacelike hypersurfaces of constant time define the slicing within the four-dimensional spacetime. Each spacelike threading corresponds to a homogeneous universe at a given epoch,  with the slicing being orthogonal to these hypersurfaces, offering a natural arrangement conducive to the definition of constant physical quantities such as density, temperature, and the speed of light on each spacelike hypersurface~\cite{Islam01,Narlikar02,Hobson06,Roos15}.Thus, our choice of coordinates emerges organically, rendering alternative considerations unnecessary. Furthermore, the derivation of the Ricci tensors and Ricci scalar curvature from the provided metric in Eq.~\eqref{ds2} further enriches the understanding of the underlying spacetime dynamics and its mathematical representation~\cite{Lee:2020zts}.Specifically, the expressions for $R_{00}$ and $R_{ii}$ are given by
\begin{align}
& R_{00} = - \frac{3}{c^2} \left( \frac{\ddot{a}}{a} - H^2 \frac{d \ln c}{d \ln a} \right)  \quad , \quad
R_{ii} = \frac{g_{ii}}{c^2} \left( 2 \frac{\dot{a}^2}{a^2} + \frac{\ddot{a}}{a} + 2 k \frac{c^2}{a^2} - H^2 \frac{d \ln c}{d \ln a} \right) \label{tR00mp} \,, \\
&R = \frac{6}{c^2} \left( \frac{\ddot{a}}{a} + \frac{\dot{a}^2}{a^2} + k \frac{c^2}{a^2} - H^2 \frac{d \ln c}{d \ln a} \right) \label{tRmp} \,.
\end{align} 
In cosmology, one treats matter as a perfect fluid, defined by its total mass density $\rho$ and isotropic pressure $P$. This density is its rest frame mass density for a perfect fluid. Within the framework of GR, the stress-energy tensor describes this perfect fluid, given by
\begin{align}
T^{\mu\nu} = \left( \rho + \frac{P}{c^2} \right) U^{\mu} U^{\nu} + P g^{\mu\nu} \label{Tmunu} \,,
\end{align}
where $U^{\mu}$ represents its four-velocity. When the fluid is in motion, a set of fundamental ({\it i.e.} comoving) observers is considered comoving with it, characterized by a four-velocity denoted as $U^{\mu} = (c, 0,0, 0)$ ~\cite{Islam01,Hobson06}. Once one establishes the metric and the stress-energy tensor, the subsequent step involves solving the Einstein Field Equations (EFEs) to elucidate the dynamics of the scale factor in the metric. These equations govern the dynamics of expansion, including the speed and acceleration of the Universe's expansion as observed between two fundamental observers. Thus, the energy-momentum tensor of the $i$-component perfect fluid with the equation of state $\omega_i$ is given by \cite{Lee:2020zts}
\begin{align}
T_{(i)\mu}^{\nu} = \text{diag} \left(-\rho_i c^2, P_i, P_i, P_i \right) \quad ,\, \text{with} \quad \rho_i c^2 = \rho_{i0} c_0^2 a^{-3(1 + \omega_i)} \label{Tmunump} \,,
\end{align}
where $c_0$ is the present value of the speed of light, $\rho_{i0}$ is the present value of mass density of the i-component, and we use $a_0 = 1$. One can derive Friedmann equations including the general DE from Eqs.~\eqref{tR00mp}-\eqref{Tmunump} 
\begin{align}
& \frac{\dot{a}^2}{a^2} + k \frac{c^2}{a^2} = \sum_{i} \frac{8 \pi G}{3} \rho_i = \frac{8 \pi G}{3} \left[ \rho_{\textrm{r}}(a)  + \rho_{\textrm{m}}(a) + \rho_{\textrm{DE}}(a) \right]  \label{tG00mp} \,, \\ 
& \frac{\dot{a}^2}{a^2} + 2 \frac{\ddot{a}}{a} +  k \frac{c^2}{a^2} - 2 H^2 \frac{d \ln c}{d \ln a}  = -8 \pi G \sum_{i} \frac{P_i}{c^2} = -8 \pi G \sum_{i} \omega_{i} \rho_{i} \nonumber \\ 
&= -8 \pi G \left[ \frac{1}{3} \rho_{\textrm{r}}(a)  + \omega_{\textrm{DE}} \rho_{\textrm{DE}}(a) \right] \label{tG11mp} \,, \\
& \frac{\ddot{a}}{a} = -\frac{4\pi G}{3} \sum_{i} \left( 1 + 3 \omega_i \right) \rho_i + H^2  \frac{d \ln c}{d \ln a}  \nonumber \\ 
&= -\frac{4\pi G}{3} \left( 2  \rho_{\textrm{r}}(a) + \rho_{\textrm{m}}(a) + \left( 1 +\omega_{\textrm{DE}} \right) \rho_{\textrm{DE}}(a) \right) +  \frac{b}{4} H^2   \label{tG11mG00mp} \,,\\
&\rho_{\textrm{r}}(a) = \rho_{\textrm{r}0} a^{-4-\frac{b}{2}} \,,  \rho_{\textrm{m}}(a) = \rho_{\textrm{m}0} a^{-3-\frac{b}{2}} \,,  \rho_{\textrm{DE}}(a) = \rho_{\textrm{DE}0} a^{-3(1+\omega_0 +\omega_a)-\frac{b}{2}} e^{-3 \omega_a (1-a)} \,, \label{rhois} \\
&\omega_{\textrm{DE}} = \omega_0 + \omega_{a} \left(1 - a \right) = \begin{cases} \omega_0 \,, \omega_{a} \neq 0  &\textrm{CPL} \\ \omega_{a} = 0  &\omega \end{cases} \label{omegaCPL} \,, 
%& \frac{\dot{a}^2}{a^2} + k \frac{c^2}{a^2}  -\frac{ \Lambda c^2}{3} = \sum_{i} \frac{8 \pi G}{3} \rho_i = \frac{8 \pi G}{3} \left[ \rho_{\textrm{r}}(a)  + \rho_{\textrm{m}}(a) + \rho_{\textrm{DE}}(a) \right]  \label{tG00mp} \,, \\ 
%& \frac{\dot{a}^2}{a^2} + 2 \frac{\ddot{a}}{a} +  k \frac{c^2}{a^2} - \Lambda c^2 - 2 H^2 \frac{d \ln c}{d \ln a}  = -8 \pi G \sum_{i} \frac{P_i}{c^2} = -8 \pi G \sum_{i} \omega_{i} \rho_{i} \nonumber \\ 
%&= -8 \pi G \left[ \frac{1}{3} \rho_{\textrm{r}}(a)  + \omega_{\textrm{DE}} \rho_{\textrm{DE}}(a) \right] \label{tG11mp} \,, \\
%& \frac{\ddot{a}}{a} = -\frac{4\pi G}{3} \sum_{i} \left( 1 + 3 \omega_i \right) \rho_i  + \frac{\Lambda c^2}{3} + H^2  \frac{d \ln c}{d \ln a}  \nonumber \\ 
%&= -\frac{4\pi G}{3} \left( 2  \rho_{\textrm{r}}(a) + \rho_{\textrm{m}}(a) + \left( 1 +\omega_{\textrm{DE}} \right) \rho_{\textrm{DE}}(a) \right) + \frac{\Lambda c^2}{3} +  \frac{b}{4} H^2   \label{tG11mG00mp} \,,\\
%&\rho_{\textrm{r}}(a) = \rho_{\textrm{r}0} a^{-4} \quad , \quad \rho_{\textrm{m}}(a) = \rho_{\textrm{m}0} a^{-3} \quad , \quad \rho_{\textrm{DE}}(a) = \rho_{\textrm{DE}0} a^{-3(1+\omega_0 +\omega_a)} e^{-3 \omega_a (1-a)} \,, \label{rhois} \\
%&\omega_{\textrm{DE}} = \omega_0 + \omega_{a} \left(1 - a \right) = \begin{cases} \omega_0 \,, \omega_{a} \neq 0  &\textrm{CPL} \\ \omega_{a} = 0  &\omega \end{cases} \label{omegaCPL} \,, \\
\end{align}
where $\rho_{\textrm{r}0}$, $\rho_{\textrm{m}0}$, and $\rho_{\textrm{DE}0}$ represent the mass-density of radiation (photon and neutrino), matter (baryon and DM), and dark energy, respectively, at the present epoch \cite{Lee:2020zts,Lee:2023bjz}.  The Chevallier-Polarski-Linder (CPL) parametrization, which assumes $\omega_{\textrm{DE}}$ to be a linear function of the scale factor $a$, is presented in Eq.~\eqref{omegaCPL} \cite{Chevallier:2000qy,Linder:2002et}. We also define the $\omega$ model when $\omega_a = 0$.

\subsection{Luminosity distance}
\label{subsec:DL}

The distance modulus, denoted by $\mu = m - M$, represents the discrepancy between the apparent magnitude $m$ (ideally corrected for interstellar absorption effects) and the absolute magnitude $M$ of an astronomical entity. It is linked to the luminosity distance $D_{\textrm{L}}$ in parsecs through the formula
\begin{align}
\mu = 5 \log_{10} \left[ \frac{D_{\textrm{L}}}{1 \textrm{Mpc}} \right] + 25 \label{mu} \,.
\end{align}
This definition proves convenient as the observed brightness of a light source correlates with its distance according to the inverse square law, and brightnesses are typically expressed in magnitudes. Absolute magnitude $M$ denotes the apparent magnitude of an object when viewed from a distance of $10$ parsecs. The relationship between magnitudes and flux $\mathcal F$ is given by
\begin{align}
m = -2.5 \log_{10} \mathcal F(D_{\text{L}})  \quad , \quad M = -2.5 \log_{10} \mathcal F(10) \label{mM} \,. 
\end{align}
The expression for E$^{(\textrm{SMC})}$ in Eq. ~\eqref{HzCC} for a flat Universe (using the CPL parametrization is obtained from Eq.~\eqref{tG00mp} 
\begin{align}
\frac{H^{(\textrm{SMC})}}{H_0} \equiv E^{(\textrm{SMC})} \simeq \sqrt{ \Omega_{\m0} a^{-3} + \left( 1 - \Omega_{\m0} \right) a^{-3(1 + \omega_0 + \omega_a)} e^{-3 \omega_a (1 -a )} } \label{EG} \,,
\end{align} 
where we ignore the radiation and the curvature contribution because we analyze the late-time Universe.
To determine the luminosity distance in the meVSL model, we need to reevaluate its fundamental definition. Here, the observed luminosity $L_0$ detected at the present epoch differs from the absolute luminosity $L_s$ of the source emitted at redshift $z$. Conservation of flux from the source to the observed point is 
\begin{align}
\mathcal F = \frac{L_s}{4 \pi D_{\textrm{L}}^2(z)} = \frac{L_0}{4 \pi D_{\textrm{M}}^2(z_0)} \label{mathF} \,.
\end{align}
The absolute luminosity, $L_s \equiv \Delta E_1/ \Delta t_1$, represents the ratio of the emitted light energy $\Delta E_1$ to the emission time interval $\Delta t_1$. Similarly, one can denote the observed luminosity as $L_0 = \Delta E_0/ \Delta t_0$. Consequently, one can rewrite the luminosity distance using Eq.~\eqref{mathF} as \cite{Lee:2020zts,Lee:2023bjz}
\begin{align}
D_{\textrm{L}}^2 (z) = \frac{L_s}{L_0} D_{\TM}^2 (z_0) = \frac{\Delta E_1}{\Delta E_0} \frac{\Delta t_0}{\Delta t_1} D_{\TM}^2 (z_0) = \left(1 + z \right)^{2-\frac{b}{4}} D_{\textrm{M}}^2 (z_0) \,, \label{DL2app}
\end{align}
where we employ
\begin{align}
\frac{\Delta E_1}{\Delta E_0} &= \frac{h_1 \nu_1}{h_0 \nu_0} = \frac{\nu_1^{(\SMC)}}{\nu_0^{(\SMC)}} = (1+z) \quad , \quad
\frac{\Delta t_0}{\Delta t_1} = \frac{\nu_1}{\nu_0} = \frac{\nu_1^{(\SMC)} (1+z)^{-b/4}}{\nu_0^{(\SMC)}} = (1+z)^{1-\frac{b}{4}} \label{t0ot1} \,,
\end{align}
where we use the cosmic evolution relation for the Planck constant $h_{i} = h_0 a_i^{-b/4}$. In the meVSL model, an expanding Universe must adhere to adiabatic expansion, leading to the cosmological evolution of the Planck constant~\cite{Lee:2022heb}. The first law of thermodynamics, which ensures energy conservation, requires that,  the entropy of the Universe remains unchanged. Consequently, the Planck constant should evolve as $h(a_i) = h_0 a_i^{-b/4}$ in this model~\cite{Lee:2020zts,Lee:2023bjz,Lee:2022heb}.
This relation also holds for the angular diameter distance $D_{\textrm{A}}$. Consequently, the luminosity distance in the meVSL model is given by \cite{Lee:2020zts}
\begin{align}
D_{\textrm{L}}(z) = \left( 1 + z \right)^{1 - \frac{b}{8}} D_{\textrm{M}}(z) = \left( 1 + z \right)^{2 - \frac{b}{8}} D_{\textrm{A}}(z) \label{CDDRmeVSL} \,.
\end{align}
Under this premise, the modification of the absolute magnitude of SNe Ia is expressed as
\begin{align}
M - M_{0} = - 2.5 \log \left[ \frac{L}{L_0} \right] = \frac{5}{4} b \log \left[ a\right] \label{MmM0} \,,
\end{align} where the subscript $0$ denotes the local value of $M$.  
The last equality of Eq.~\eqref{MmM0} can be obtained from the following~\cite{Lee:2020zts}. SNe Ia are nuclear explosions of white dwarfs (WDs) in binary systems, where the WD accretes matter from a companion until it approaches the Chandrasekhar limit. This limit represents the maximum mass a WD can have before electron degeneracy pressure fails to counteract gravitational collapse. For WDs, this limit is typically around $1.4$ solar masses. If a WD exceeds this mass, it can collapse into a neutron star or black hole, while those below remain stable.

The Chandrasekhar mass limit, M$_{\textrm{Ch}}$, is determined by the equation of state for an ideal Fermi gas, with a constant $\omega_0^3 \approx 2.018$ (a constant related to the solution for the so-called Land-Emden equation), the average molecular weight per electron $\mu_e$, and the mass of the hydrogen atom $m_\textrm{H}$ ~\cite{Lee:2020zts}
\begin{align} \textrm{M}_{\textrm{Ch}} = \frac{\omega_{0}^3 \sqrt{3\pi}}{2} \left( \frac{\hbar c}{G} \right)^{\frac{3}{2}} \frac{1}{\left( \mu_{e} m_{\textrm{H}}\right)^2} \equiv  \textrm{M}_{\textrm{Ch} 0} a^{-\frac{b}{2}} \label{MCh} \,.
\end{align}
The peak luminosity of SNe Ia is proportional to the mass of synthesized nickel, which is a fraction of the Chandrasekhar mass. Consequently, the absolute magnitude of SNe Ia, which measures luminosity, is related to the Chandrasekhar mass and the total amount of nickel synthesized $L \propto \textrm{M}_{\textrm{Ch}} \propto a^{-\frac{b}{2}}$. The absolute magnitude $M$ is given by $M \propto -2.5 \log [L] \propto \frac{5}{4}b \log[a]$. Thus, the distance modulus of meVSL, $\mu \equiv m - M$, is written as:
\begin{align}
\mu(z) &= m - M = 5 \log_{10} \left[ \frac{D_{\TL}}{\TMpc} \right] + 25 \quad , \quad D_{\TL} = \frac{c_0}{H_0} d_{\TL}(z) \,\,, \,\, d_{\TL}(z) \equiv \int_{0}^{z} \frac{dz'}{E^{(\SMC)}(z')} \label{mumeVSL} \,,
\end{align}
where $E^{(\SMC)}$ is in Eq.~\eqref{EG}. The theoretically predicted apparent magnitude $m_{\textrm{th}}$ can be obtained from Eqs.~\eqref{MmM0} and \eqref{mumeVSL}
\begin{align}
m_{\Tth}(z) &= M_0 + 5 \log \left[ \frac{D_{\TL}}{1 \TMpc} \right] + 25 + \frac{5}{4} b \log \left[ a\right] \nonumber \\ 
&= M_0 + 42.3841 - 5 \log \left[\textrm{h} \right] - \frac{15}{8} b \log  \left[ (1+z) \right] + 5 \log \left[ d_{\TL} \right] \nonumber \\
&\equiv \mathcal M + 5 \log \left[ d_{\TL} \right] \label{muthmeVSL} \,.
\end{align}

\section{Statistical Analysis}
\label{sec:chi2}

We examine constraints on the evidence for cosmic variation in the speed of light using the Pantheon$+$ dataset~\cite{Scolnic:2017caz}. The Pantheon compilation includes $1048$ SNe Ia covering a redshift range $0.01 \leq z \leq 2.3$. This dataset incorporates $365$ spectroscopically confirmed SNe Ia from the Pan-STARRS1 (PS1) Medium Deep Survey, along with a subset of $279$ PS1 SNe Ia (with redshifts ranging from $0.03$ to $0.68$) with reliable distance estimates derived from various sources, including SDSS, SNLS, and HST samples.  To determine cosmological parameters using $H(z)$, higher-redshift SNe Ia are employed, and the degenerate parameters $\mathcal M$ are typically marginalized as nuisance parameters \cite{SDSS:2014iwm,Pan-STARRS1:2017jku}.  For example, Eq.~\eqref{muthmeVSL} is used to construct and minimize $\bar{\chi}^2(\Omega_{\m0}) \equiv \int d \mathcal M \chi^2(\mathcal M\,, \Omega_{\m0})$ where the degenerate combination provided in the same equation \cite{Arjona:2018jhh,Kazantzidis:2020tko}.  However, marginalizing the parameter $\mathcal M$ can result in the loss of valuable physical information regarding potential spatial variations of $H_0$ and/or temporal variations of the absolute magnitude $\mathcal M$. For instance, an  $\mathcal M$ value that evolves with redshift, resulting in low $\mathcal M$ values at low $z$, could indicate either higher local values of $H_0$ due to a local matter underdensity or lower values of the absolute magnitude  $\mathcal M$ in recent cosmological times, possibly caused by a time variation of Newton’s constant. Specifically, since we are studying the potential variation of the speed of light with cosmic time, we do not marginalize other parameters in this manuscript when we use the minimal $\chi^2$ method \cite{Shanks:2018rka,Shanks:2019inu,Lukovic:2019ryg,Bohringer:2019tyj}. The chi-squared ($\chi^2$) represents a weighted summation of squared deviations, given by 
\begin{align}
\chi^2 = \sum_{i, j} \left( m_{i,\Tobs} - m_{i, \Tth} \right) C^{-1}_{ij} \left( m_{i,\Tobs}- m_{j, \Tth} \right) \label{chi2} \,,
\end{align} where $m_{i,\textrm{obs}}$ signifies the observed apparent magnitude, $m_{i,\textrm{th}}$ represents the theoretical apparent magnitude of SNe Ia at the redshift $z_i$ as defined in Eq.~\eqref{mumeVSL}, and $C_{ij} = D_{ij} + C_{\textrm{sys}}$ denotes the covariance matrix. Here, $D_{ij} = \sigma_i^2 \delta_{ij}$ stands for the variance of each observation, and $C_{\textrm{sys}}$ is a non-diagonal matrix associated with systematic uncertainties. $m_{\Tth}$ is a function of $M\,,\textrm{h}\,,\Omega_{\m0}\,,\omega_0\,,$ and $\omega_{a}$. The reduced chi-square statistic defined as chi-square per degree of freedom is used extensively in the goodness of fit testing 
\begin{align}
\chi^2_{\nu} = \frac{\chi^2}{\nu} \label{chinu2} \,,
\end{align}
where the degree of freedom $\nu = N - p$, signifies the number of observations $N$ minus the number of fitted parameters $p$. As a heuristic, when the variance of the measurement error is known a priori, a $\chi^2_{\nu} \gg 1$ suggests a substandard model fit. Conversely, a $\chi^2_{\nu} > 1$ implies that the fit has not adequately captured the data (or that the error variance has been underestimated). Ideally, a $\chi^2_{\nu}$ around $1$ indicates that the correspondence between observations and estimates is congruent with the error variance.  Since all models have nearly identical reduced chi-square values, this implies that all models are statistically equally viable. 

%%%%%%%%%%%%%%%%%%%%%%%%%%%%%%%%%%%%%%%%%%
\section{Bounds for the variation of $c$ for different models}
\label{sec:dotc}

The previous constraint on the temporal variation of the speed of light $c$ was derived from the variation in the radius of a planet~\cite{Racker:2007hj}. However, this constraint stemmed from the analysis of the time-varying radius of Mercury~\cite{McElhinny:1978na} using a specific model known as the covariant variable speed of light theory proposed by Magueijo~\cite{Magueijo:2000zt}. Consequently, this constraint cannot be directly applied within the framework of the meVSL model.

To explore the variation of the speed of light over time, we turn to SNe Ia, which serve as reliable standard candles for probing the cosmic expansion rate in the late Universe. Our investigation focuses on utilizing data from the Pantheon SNe Ia catalog~\cite{Scolnic:2017caz}. Specifically, we delve into two primary models: the $\omega$CDM model, which assumes a constant $\omega$, and the CPL dark energy model. Through this analysis, we aim to shed light on the potential temporal evolution of the speed of light and its implications within these cosmological frameworks.

First, in the first row of Table~\ref{tab:omegaw0}, we considered the $\Lambda$CDM models ($\omega_0 = -1, \omega_a = 0$, and $b=0$). When varying the local absolute magnitude $M_0$ from -19.35 to -19.55, we found that the best-fit value of $\Omega_{\m0}$ remains unchanged at 0.285, while the best-fit values of h decrease from $0.702$ to $0.640$.  For this model, the 1-$\sigma$ values for $M_0$, $\Omega_{\m0}$, and $h$ are $(-19.6512, -19.0732)$, $(0.273, 0.297)$, and $(0.605, 0.791)$, respectively. This suggests the interesting possibility that the h value derived from SNe Ia data can be similar to that of Planck data, potentially alleviating the Hubble tension. If we also allow the value of $M_0$ to vary, its best-fit value becomes $-19.36$, and the best-fit value of h is $0.698$. Similar results were discussed in the reference \cite{Perivolaropoulos:2021bds}.  $\Lambda$CDM refers to the case where we exclude VSL models. We present this model solely for comparison purposes. In our manuscript, we investigate the meVSL model for $\omega$CDM and CPL models.

The main interest of this manuscript is whether the speed of light can vary in models with values of cosmological parameters similar to those of the SMC. Therefore, we will limit our discussions to such models. Among the CPL models, there are cases where the dark matter density is zero. There are also interesting papers interpreting these results \cite{Gueorguiev:2022wit,Gupta:2024eqo}. The so-called CCC (covarying coupling constants) + TL (tired light) cosmology obtains $\Omega_{\m0} = 0$ using the baryonic acoustic oscillation data \cite{Gupta:2024eqo}.  One can find the origin of dark matter and dark energy as a time-dependent conformal scale factor in the Scale Invariant Vacuum (SIV) paradigm \cite{Gueorguiev:2022wit}. However, we will treat this outcome as merely a result of statistical analysis and will not consider any further interpretation or implication.  Although this could be an interesting topic for another paper, as mentioned earlier, in this manuscript, we will focus primarily on models where the speed of light can vary with cosmological parameters consistent with the SMC. Also,  as evident from equations \eqref{EG} and \eqref{muthmeVSL}, we acknowledge the degeneracy relationship between $\Omega_{\m0}$ and the $\omega_a$ term in the exponential exponent. Thus, we accept that the value of $\Omega_{\m0}$ can vary due to this term, and we refrain from further physical interpretations or alternative explanations.

\section{$c$ for $\omega$CDM }
\label{sec:wCDM}

We explore the $\omega$CDM models, $\omega_{a} = 0$, in Eq.~\eqref{omegaCPL}. Utilizing a maximum likelihood analysis, we examine various models characterized by varying cosmological parameters. We show the results of these analyses in Table~\ref{tab:omegaw0}, which reveals intriguing insights into the relationships between these parameters.

Within this table, we uncover significant patterns and correlations among cosmological parameters. Notably, when $\omega_{0}$ is held constant, both $M_0$ and $\textrm{h}$ exhibit degeneracy, as do $\Omega_{\m0}$ and $b$. Consequently, fixing $\Omega_{\m0}$ results in nearly identical values for $M_0$ and h, with only $b$ values varying. Conversely, when we fix the value of $\textrm{h}$, only the $M_0$ values change, while $\Omega_{\m0}$ and $b$ become irrelevant in this context.

%\textcolor{blue}{First, in the first row of Table~\ref{tab:omegaw0}, we considered the $\Lambda$CDM models (\textit{i.e.}, $\omega_0 = -1, \omega_a = 0$, and $b=0$). When varying the absolute brightness from $-19.35$ to $-19.55$, we found that the best-fit value of $\Omega_{\m0}$ remains unchanged at $0.285$, while the best-fit values of h decrease from $0.702$ to $0.640$. This suggests the interesting possibility that the h value derived from SNe Ia data can be similar to that of Planck data, potentially alleviating the Hubble tension. If we also allow the value of $M_0$ to vary, its best-fit value becomes $-19.36$, and the best-fit value of h is $0.698$. Similar results were discussed in the reference. \cite{Perivolaropoulos:2021bds}.}

Furthermore, we observe that among the $\omega$CDM models derived from the Pantheon$+$ data, those displaying noticeable time variations in the speed of light are specifically characterized by $\omega_{0} = -1$ and $\Omega_{\m0} \geq 0.30$. This observation highlights the complex interplay between cosmological parameters and their implications for understanding the temporal evolution of fundamental physical constants.

 \begin{table}[h!]
		\caption{Best fit values and their corresponding 1-$\sigma$ uncertainties for both $\Lambda$CDM ($\omega_{a} = 0$ and $b=0$) and $\omega$CDM models (with $\omega_{a} = 0$) are presented. Models highlighted in green indicate potential meVSL models for $\omega$CDM.}
		\label{tab:omegaw0}
\footnotesize
\begin{adjustwidth*}{}{}
 	\begin{center}
%\centering %% If there is a figure in wide page, please release command \centering
		\begin{tabular}{|c|c|c|c|c|c|c|c|c|} 
			\hline
			 Models& Submodels & $M_0$ & $\omega_0$ & $\Omega_{m0} $ & $h$  &  $b$ & $\nu$ & $\chi_{\nu}^2$ \\ \hline
			 %\multirow{8}{*}{Equivalence} & redshift $z$ & speed of light $\tc = \tc_0 a^{\frac{b}{4}}$ \\
			  \multirow{4}{*}{$\Lambda$CDM} & \multirow{3}{*}{fixing $M_0$ } &$-19.3500$  & $-1$ & $0.285 \pm 0.012$ &$0.702 \pm 0.002 $ &$0$ & $1046$ & $0.988$ \\  
			  & &$-19.4500$  & $-1$ & $0.285 \pm 0.012$ &$0.670 \pm 0.002 $ &$0$ & $1046$ & $0.989$ \\ 
			  & &$-19.5500$  & $-1$ & $0.285 \pm 0.012$ &$0.640 \pm 0.002 $ &$0$ & $1046$ & $0.989$ \\ \cline{2-9}
			  & &$-19.3622 \pm 0.2890$  & $-1$ & $0.285 \pm 0.012$ &$0.698 \pm 0.093 $ &$0$ & $1045$ & $0.989$ \\ \hline 
			  \multirow{6}{*}{$\omega_0 = -1$} &\multirow{4}{*}{fixing $\Omega_{\m0}$ } &$-19.3556 \pm 0.2899$  & $-1$ & $0.28$ &$0.700 \pm 0.093 $ &$0.009 \pm 0.022$ & $1045$ & $0.989$ \\
			  &  &$-19.3558 \pm 0.2899$  & $-1$ & $0.29$ &$0.700 \pm 0.093 $ &$-0.009 \pm 0.022$ & $1045$ & $0.989$ \\
			  & &$-19.3561 \pm 0.2899$  & $-1$ & $0.30$ &$0.700 \pm 0.093 $ &\cellcolor{green}$-0.027 \pm 0.022$ & $1045$ & $0.989$ \\ 
			  & &$-19.3563 \pm 0.2899$  & $-1$ & $0.31$ &$0.700 \pm 0.093 $ &\cellcolor{green}$-0.044 \pm 0.022$ & $1045$ & $0.989$ \\
			  & \ref{subsec:w0m1fOm0} &$-19.3566 \pm 0.2898$  & $-1$ & $0.32$ &$0.700 \pm 0.093 $ &\cellcolor{green}$-0.061 \pm 0.022$ & $1045$ & $0.989$ \\ \cline{2-9}
			  & fixing h &$-19.4395 \pm 0.0072$  & $-1$ & $0.299 \pm 0.111$ &$0.6736$ &$-0.025 \pm 0.193$ & $1045$ & $0.989$ \\ %\multirow{2}{*}{fixing h}
			  & \ref{subsec:w0m1fh} &$-19.2353 \pm 0.0072$  & $-1$ & $0.299 \pm 0.111$ &$0.74$ &$-0.025 \pm 0.193$ & $1045$ & $0.989$ \\ \hline
			  \multirow{8}{*}{$\omega$CDM} & fixing h &$-19.4525 \pm 0.0071$  & $-1.23 \pm 0.05$ & $0.380 \pm 0.085$ &$0.6736$ &$-0.057 \pm 0.159$ & $1044$ & $0.988$ \\
			  & \ref{subsec:w0fh} &$-19.2481 \pm 0.0071$  & $-1.22 \pm 0.05$ & $0.378 \pm 0.087$ &$0.74$ &$-0.055 \pm 0.163$ & $1044$ & $0.988$ \\ \cline{2-9} % \multirow{2}{*}{fixing $h$} 
			   & \multirow{4}{*}{fixing $\omega_{0}$} &$-19.3655 \pm 0.2893$  & $-0.9$ & $0.311 \pm 0.106$ &$0.695 \pm 0.093 $ &$-0.106 \pm 0.163$ & $1044$ & $0.993$ \\
			  & &$-19.3659 \pm 0.2847$  & $-0.95$ & $0.301 \pm 0.092$ &$0.696 \pm 0.091 $ &$-0.058 \pm 0.152$ & $1044$ & $0.992$ \\
			  & &$-19.3728 \pm 0.2877$  & $-1.0$ & $0.299 \pm 0.111$ &$0.695 \pm 0.092 $ &$-0.025 \pm 0.193$ & $1044$ & $0.990$ \\
			  & &$-19.3676 \pm 0.2908$  & $-1.05$ & $0.290 \pm 0.110$ &$0.697 \pm 0.093 $ &$0.023 \pm 0.305$ & $1044$ & $0.989$ \\
			  & \ref{subsec:w0fw0}  &$-19.3642 \pm 0.2895$  & $-1.1$ & $0.288 \pm 0.134$ &$0.699 \pm 0.093 $ &$0.054 \pm 0.262$ & $1044$ & $0.988$ \\
			\cline{2-9}
			& No fixing & $-19.5513 \pm 0.1284$  & $-1.22 \pm 0.04$ & $0.335 \pm 0.088$ &$0.644 \pm 0.038 $ &$0.022 \pm 0.176$ & $1043$ & $0.989$ \\
			\hline
		\end{tabular}
	\end{center}
\end{adjustwidth*}
 \end{table}

\subsection{$\omega_{0} = -1$ with fixing $\Omega_{\m0}$}
\label{subsec:w0m1fOm0} 

We keep $\Omega_{\m0}$ constant and perform a $\chi^2$ test on these models. Notably, the value of h remains unchanged even as we vary $\Omega_{\m0}$ within the range of $0.28$ to $0.32$. As $\Omega_{\m0}$ increases, both $M_0$ and $b$ decrease. Especially, the best-fit value of $b$ is positive only for $\Omega_{\m0} = 0.28$, while for other values within this range, the best-fit values of $b$ are negative. 

\begin{figure*}
\centering
\vspace{1cm}
\begin{tabular}{cc}
\includegraphics[width=0.5\linewidth]{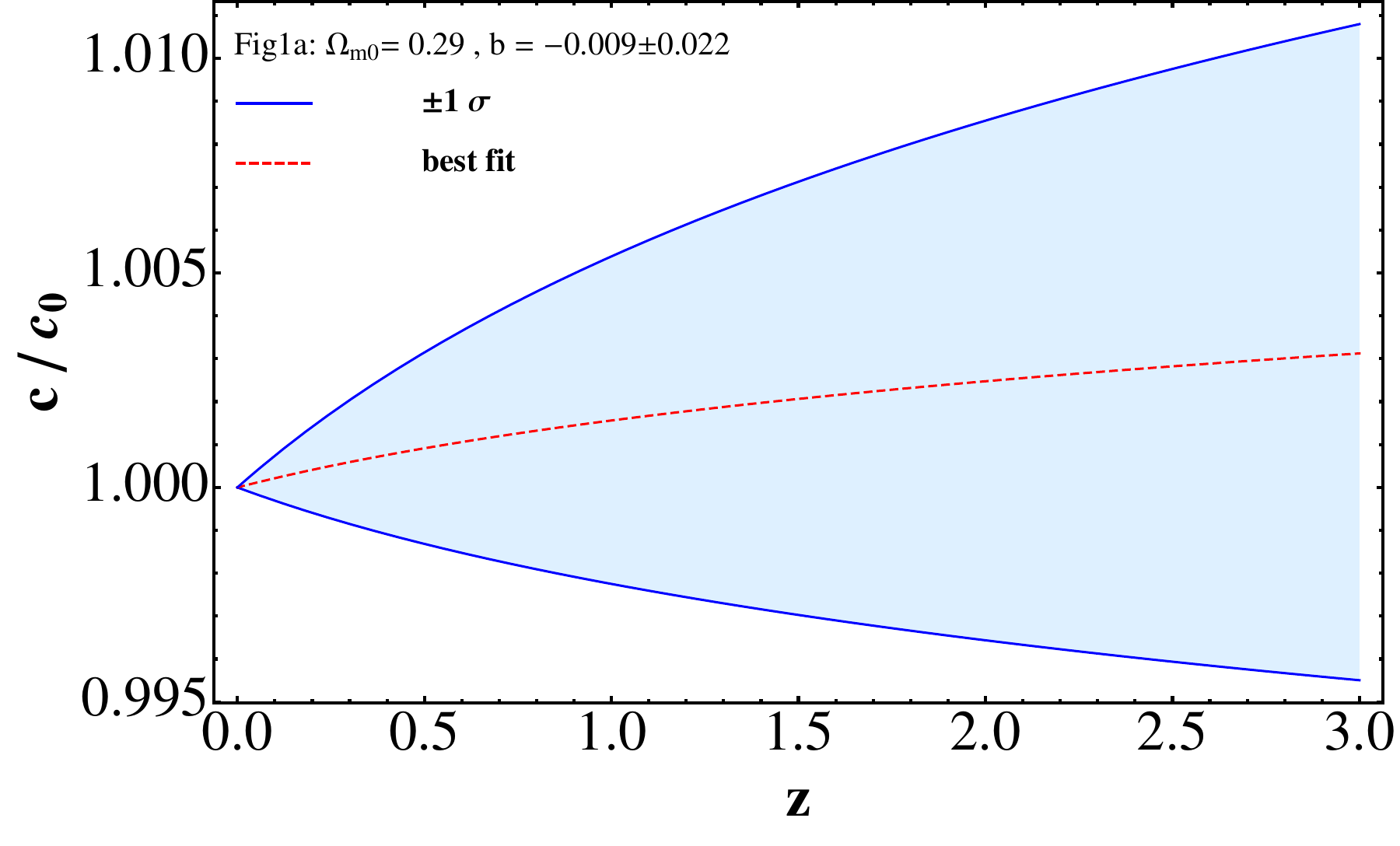} &
\includegraphics[width=0.49\linewidth]{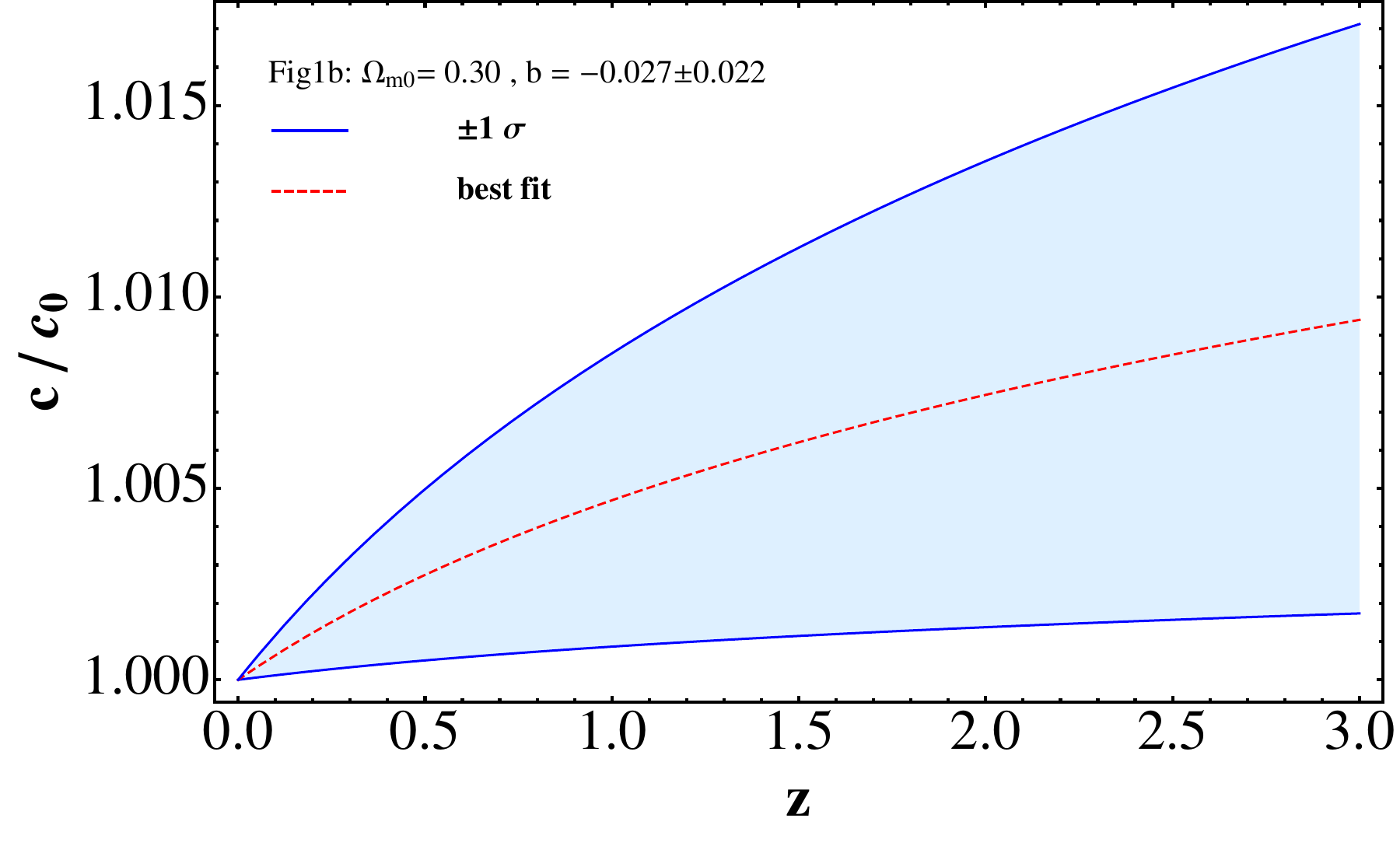}
\end{tabular}
\vspace{-0.5cm}
\caption{The plots depict the ratios of $c$ to its present value, $c_0$, as a function of redshift for different $\Omega_{\Tm0}$ values, with $\omega_0 = -1$. In panel (a), we observe $c/c_0$ for $\Omega_{\m0} = 0.29$, where the dashed line represents the best-fit value, while the solid lines denote the $1$-$\sigma$ error margins. Panel (b) displays $c(z)/c_0$ for $\Omega_{\m0} = 0.30$, with the dashed line indicating the best-fit value and the solid lines indicating the 1-$\sigma$ errors. } \label{fig-1}
\vspace{1cm}
\end{figure*}

Within the 1-$\sigma$ error range, the values of $b$ show both positive and negative trends for $\Omega_{\m0} = 0.28$ and $0.29$. However, for $0.30 \leq \Omega_{\m0} \leq 0.32$, the 1-$\sigma$ region consistently yields negative values for $b$, indicating a decrease in the speed of light over cosmic time. These trends are illustrated in Fig.~\ref{fig-1}.

In panel a of Fig.~\ref{fig-1}, we depict the cosmic evolution of the best-fit value of $c/c_0$ along with its 1-$sigma$ errors for the model with $\omega_0 = -1$ and $\Omega_{\m0} = 0.29$. Here, the best-fit value of $b$ is $-0.009$, suggesting an increase in the speed of light with increasing $z$. However, the uncertainty in $b$ allows for both negative and positive values within the 1-$\sigma$ error range, leading to ambiguity regarding the variation in the speed of light for this model.

For $\Omega_{\m0} \geq 0.30$, the best-fit value of $b$ is consistently negative within the 1-$\sigma$ error range, indicating a monotonically decreasing speed of light over cosmic time. In panel b of Fig.~\ref{fig-1}, we show the cosmological evolution of $c/c_0$ for the model with $\omega_0 = -1$ and $\Omega_{\m0} = 0.30$. In this case, the best-fit value of $b$ is $-0.027$, further supporting the decrease in the speed of light over time.

Additionally, the ratio of the time variation of the speed of light to its present value, expressed as $\dot{c}_0 / c_0 = \frac{b}{4} H_0$, is constrained within the 1-$\sigma$ ranges of $(-8.76, -0.89)$, $(-11.80, -3.93)$, and $(-14.84, -6.98)$ for $\Omega_{\m0} = 0.30$, 0.31, and 0.32, respectively. These constraints represent significant improvements over those reported in \cite{Racker:2007hj}.

\subsection{$\omega_{0} = -1$ without fixing $\Omega_{\m0}$}
\label{subsec:w0m1vOm0} 

Within the 1-$\sigma$ error range, the cosmological parameters for the model with $\omega_{0} = -1$ are $-19.6605 \leq M_0 \leq -19.0851$, $0.188 \leq \Omega_{\m0} \leq 0.410$, $0.603 \leq h \leq 0.787$, and $-0.218 \leq b \leq 0.168$. Compared to the $\Lambda$CDM model, the constraint on $\Omega_{\m0}$ in the meVSL model is significantly weaker. It is because, as can be understood from equation~\eqref{muthmeVSL}, both h and $b$ contribute to $m_{\Tth}$, allowing the effect of changes in h to yield similar results from changes in $b$.

In the meVSL framework, the cosmological evolution of the speed of light is described by $c = c_0 (1+z)^{-b/4}$. The best-fit value of $b$ is $-0.025$, suggesting that the speed of light was higher in the past than it is today. However, the 1-$\sigma$ range of $b$ includes both negative and positive values, making it inconclusive to definitively determine the variation of the speed of light in this model.

%%%%%%%%%%%%%%%%%%%%%%%%%%%%%%%%%%%%%%
%%%%%%%%%%%%%%%%%%%%%%%%%%%%%%%%%%%%%%%

\subsection{$\omega_{0} = -1$ with fixing h}
\label{subsec:w0m1fh} 

 We keep the value of h fixed and perform a maximum likelihood analysis. In these scenarios, both $\Omega_{\m0} (= 0.299 \pm 0.111)$ and $b ( = -0.025 \pm 0.193)$ remain stable despite variations in h ranging from  $0.6376$ to $0.74$. As h increases, $M_0$ also experiences a corresponding increase. The best-fit value of $b$ remains consistent at $ -0.025$ across these models, indicating a consistent decreasing trend in the speed of light over cosmic time. We also understand from Equation \eqref{muthmeVSL} that the variation in $M_0$ is insensitive to changes in h. However, within the 1-$\sigma$ error margin, the values of $b$ exhibit both positive and negative trends. Consequently, it is plausible to conclude that there is no significant variation in the speed of light within these models.

\subsection{Fixing h}
\label{subsec:w0fh} 

We perform a maximum likelihood analysis while varying other parameters under a fixed value of h. Under this condition, while we systematically vary h from $0.6736$ to $0.74$, the best-fit values of parameters $M_0$, $\omega_0$, and $b$ all show an increasing trend. However, concurrently, the best-fit value of $\Omega_{\m0}$ exhibits a decrease as h increases.  Furthermore, as h increases, $M_0$ also shows an upward trend. Notably, the best-fit values of $b$ consistently remain negative across these models, showing a consistent decrease in the speed of light over cosmic time.  Nonetheless, within the $1$-$\sigma$ error range, the values of $b$ demonstrate a variability, with some values being positive and others negative, suggesting no clear discernible pattern in the variation of the speed of light for these models. 

\subsection{Fixing $\omega_{0}$}
\label{subsec:w0fw0} 

We conduct a maximum likelihood analysis for $\omega$CDM models, with no fixed cosmological parameters except $\omega_0$. The range of $\omega_0$ is varied from $-0.9$ to $-1.1$. As $\omega_0$ decreases, $\Omega_{\m0}$ decreases while $b$ increases. However, consistent trends were not observed in the changes of $M_0$ and h across these models. For models with $\omega_0 \geq -1.0$, the best-fit values of $b$ were negative, while positive best-fit values of $b$ were obtained for models with $\omega_0 \leq -1.0$. Nevertheless, within the $1$-$\sigma$ error range, the values of $b$ for all models included both positive and negative values, suggesting no clear evidence of variations in the speed of light in these models.

%We conduct a maximum likelihood analysis for $\omega$CDM models without fixing any cosmological parameters except $\omega_0$. We vary the value of $\omega_0$ from $-0.9$ to $-1.1$. As $\omega_0$ decreases, $\Omega_{\m0}$ also decreases while $b$ increases. However, there are no consistent trends observed in the changes of $M_0$ and h across these models. For models with $\omega_0 \geq -1.0$, the best-fit values of $b$ are negative. Conversely, we obtain positive best-fit values of $b$ for models with $\omega_0 \leq -1.0$. Nonetheless, within the $1$-$\sigma$ error range, the values of $b$ for all models include both positive and negative values, suggesting no clear evidence of variations in the speed of light in these models.

\subsection{WIthout fixing}
\label{subsec:w0} 

Finally, we conduct a maximum likelihood analysis without fixing any parameters. In this case, the best-fit value of $M_0$ decreases with increasing $\Omega_{\m0}$, reaching $0.335$, while the values of $\omega_0$ and $h$ decrease to $-1.22$ and $0.644$, respectively. These are shown in the last row of Table~\ref{tab:omegaw0}.

\section{$c$ for CPL }
\label{sec:CPL}

In this section, we perform a maximum likelihood analysis for the CPL models, exploring various scenarios across different cosmological parameter values. The results of this analysis are summarized in Table~\ref{tab:CPL}. Initially, we examine the case of GR, represented by $b = 0$, and then extend the analysis to include the meVSL models, where $b$ is allowed to vary. Unlike the $\omega$CDM models, obtaining viable values for $\Omega_{\m0}$ from this analysis without any prior constraints proves challenging. Therefore, we choose to exclusively perform the maximum likelihood analysis for fixed values of $\Omega_{\m0}$.

 \begin{table}[h!]
		\caption{Presented below are the best-fit values and their corresponding 1-$\sigma$ errors for cosmological parameters in CPL models. Only models highlighted in green (or cyan) indicate the potential for exhibiting time variations in the speed of light. The $b$ values highlighted in green indicate negative values, and those in cyan indicate positive values. These represent a clear monotonic decrease or increase in the speed of light with cosmic time, respectively.}
		\label{tab:CPL}

\begin{adjustwidth*}{}{}
\footnotesize
 	\begin{center}
%\centering %% If there is a figure in wide page, please release command \centering
		\begin{tabular}{|c|c|c|c|c|c|c|c|c|} 
			\hline
			 Models& $M_0$ & $\omega_0$ & $\omega_{a}$ & $\Omega_{m0} $ & $h$  &  $b$ & $\nu$ & $\chi_{\nu}^2$ \\ \hline
			 %\multirow{8}{*}{Equivalence} & redshift $z$ & speed of light $\tc = \tc_0 a^{\frac{b}{4}}$ \\
			  \multirow{2}{*}{$b = 0$} &$-19.3617 \pm 0.2914$  & $-1$ & $1.36 \pm 0.42$ & $0.163 \pm 0.054$ &$0.704 \pm 0.094 $ &$0$ & $1044$ & $0.987$ \\ %\hdashline
			                        &$-19.3582 \pm 0.2915$  & $-1.22 \pm 0.15$ & $0$ & $0.348 \pm 0.036$ &$0.704 \pm 0.094 $ &$0$ & $1044$ & $0.988$ \\ 
			                      \ref{subsec:b0}  &$-19.3325 \pm 0.0908$  & $-0.76 \pm 0.02$ & $1.30 \pm 0.25$ & $-0.062 \pm 0.068$ &$0.712 \pm 0.030 $ &$0$ & $1043$ & $0.988$ \\ \hline
			    $b \neq 0$ &$-19.2231 \pm 0.1985$  & $-1$ & $1.44 \pm 0.08$ & $-0.102 \pm 0.065$ &$0.748 \pm 0.068 $ &$0.464 \pm 0.156$ & $1043$ & $0.988$ \\ %\hdashline  \multirow{2}{*}{$b \neq 0$} 
			       \ref{subsec:w0m1wa0}        &$-19.5513 \pm 0.1284$  & $-1.22 \pm 0.04$ & $0$ & $0.335 \pm 0.088$ &$0.644 \pm 0.038 $ &$0.022 \pm 0.176$ & $1043$ & $0.989$ \\ \hline                    
			   \multirow{4}{*}{$\omega_0 = -0.95$} &$-19.4115 \pm 0.2313$  & $ $ & $1.85 \pm 0.22$ & $0.28$ &$0.688 \pm 0.073 $ &$\cellcolor{green}-0.273 \pm 0.014$ & $1044$ & $0.987$ \\
			  &$-19.4065 \pm 0.2324$  & $ $ & $1.90 \pm 0.23$ & $0.29$ &$0.690 \pm 0.074 $ &\cellcolor{green}$-0.291 \pm 0.014$ & $1044$ & $0.987$ \\
			  &$-19.4110 \pm 0.2328$  & $-0.95$ & $1.95 \pm 0.23$ & $0.30$ &$0.689 \pm 0.074 $ &\cellcolor{green}$-0.309 \pm 0.014$ & $1044$ & $0.987$ \\
			  &$-19.4114 \pm 0.2335$  & $$ & $2.01 \pm 0.24$& $0.31$ & $0.689 \pm 0.074 $ &\cellcolor{green}$-0.326 \pm 0.014$ & $1044$ & $0.987$ \\
			\ref{subsec:w0m95}  &$-19.4119 \pm 0.2342$  & $$ & $2.07 \pm 0.24$ & $0.32$ &$0.688 \pm 0.074 $ &\cellcolor{green}$-0.344 \pm 0.014$ & $1044$ & $0.987$ \\ \hline
			  \multirow{4}{*}{$\omega_0 = -1.0$} &$-19.4108 \pm 0.2322$  & $$ & $1.73 \pm 0.23$ & $0.28$ &$0.689 \pm 0.074 $ &$\cellcolor{green}-0.215 \pm 0.015$ & $1044$ & $0.988$ \\
			  &$-19.3848 \pm 0.2349$  & $$ & $1.79 \pm 0.24$ & $0.29$ &$0.697 \pm 0.075 $ &\cellcolor{green}$-0.235 \pm 0.015$ & $1044$ & $0.988$ \\
			  &$-19.4036 \pm 0.2344$  & $-1.0$ & $1.85 \pm 0.24$ & $0.30$ &$0.691 \pm 0.075 $ &\cellcolor{green}$-0.255 \pm 0.015$ & $1044$ & $0.988$ \\
			  &$-19.4100 \pm 0.2346$  & $$ & $1.91 \pm 0.24$& $0.31$ & $0.689 \pm 0.074 $ &\cellcolor{green}$-0.274 \pm 0.015$ & $1044$ & $0.988$ \\
			 \ref{subsec:w0m1} &$-19.4107 \pm 0.2353$  & $$ & $1.97 \pm 0.25$ & $0.32$ &$0.689 \pm 0.075 $ &\cellcolor{green}$-0.293 \pm 0.015$ & $1044$ & $0.988$ \\ \hline
			  \multirow{4}{*}{$\omega_0 = -1.05$} &$-19.4038 \pm 0.2339$  & $$ & $1.57 \pm 0.24$ & $0.28$ &$0.691 \pm 0.074 $ &$\cellcolor{green}-0.154 \pm 0.015$ & $1044$ & $0.988$ \\
			  &$-19.4076 \pm 0.2344$  & $$ & $1.62 \pm 0.25$ & $0.29$ &$0.689 \pm 0.074 $ &\cellcolor{green}$-0.175 \pm 0.015$ & $1044$ & $0.988$ \\
			  &$-19.4079 \pm 0.2351$  & $-1.05$ & $1.68 \pm 0.25$ & $0.30$ &$0.689 \pm 0.075 $ &\cellcolor{green}$-0.195 \pm 0.015$ & $1044$ & $0.988$ \\
			  &$-19.4096 \pm 0.2356$  & $$ & $1.73 \pm 0.25$& $0.31$ & $0.689 \pm 0.075 $ &\cellcolor{green}$-0.213 \pm 0.015$ & $1044$ & $0.988$ \\
			  \ref{subsec:w0m105} &$-19.4082 \pm 0.2367$  & $$ & $1.84 \pm 0.26$ & $0.32$ &$0.690 \pm 0.075 $ &\cellcolor{green}$-0.240 \pm 0.015$ & $1044$ & $0.988$ \\ \hline
			  \multirow{2}{*}{fixing $\Omega_{m0}$} &$-19.3499 \pm 0.1176$  & $-1.18 \pm 0.04$ & $0.72 \pm 0.23$ & $0.28$ &$0.707 \pm 0.038 $ &$\cellcolor{cyan} 0.038 \pm 0.015$ & $1043$ & $0.988$ \\
			  %&$-19.4076 \pm 0.2857$  & $-1.05$ & $1.62 \pm 0.83$ & $0.29$ &$0.689 \pm 0.091 $ &\cellcolor{green}$-0.175 \pm 0.115$ & $1044$ & $0.988$ \\
			  &$-19.3465 \pm 0.1199$  & $-1.18 \pm 0.04$ & $0.76 \pm 0.25$ & $0.30$ &$0.708 \pm 0.039 $ &$0.001 \pm 0.015$ & $1043$ & $0.989$ \\
			  %&$-19.4096 \pm 0.2855$  & $-1.05$ & $1.73 \pm 0.67$& $0.31$ & $0.689 \pm 0.090 $ &\cellcolor{green}$-0.213 \pm 0.103$ & $1044$ & $0.988$ \\
			  \ref{subsec:cplOm0} &$-19.3478 \pm 0.1225$  & $-1.20 \pm 0.04$ & $0.71 \pm 0.26$ & $0.32$ &$0.708 \pm 0.040 $ &\cellcolor{green}$-0.024 \pm 0.015$ & $1043$ & $0.989$ \\ \hline
			 fixing h &$-19.4544 \pm 0.0072$  & $-0.99 \pm 0.02$ & $1.31 \pm 0.11$ & $0.033 \pm 0.061$ &$0.6736$ &$0.207 \pm 0.120$ & $1043$ & $0.988$ \\
			 \ref{subsec:cplh} &$-19.2475 \pm 0.0078$  & $-0.99 \pm 0.01$ & $1.45 \pm 0.07$ & $-0.112 \pm 0.036$ &$0.74$ &$0.460 \pm 0.094$ & $1043$ & $0.988$ \\ \hline
			  No fixing & $-19.3500 \pm 0.0857$  & $-0.76 \pm 0.02$ & $1.30 \pm 0.10$ & $-0.054 \pm 0.086$ &$0.706 \pm 0.028$ &$-0.023 \pm 0.108$ & $1042$ & $0.989$ \\ \hline
		\end{tabular}
	\end{center}
\end{adjustwidth*}
 \end{table}

\subsection{$b = 0$}
\label{subsec:b0}

We investigate CPL models within the framework of GR, initially setting $b = 0$. We analyze three scenarios: $\omega_0 = -1$, $\omega_{a} = 0$, and without fixing $\omega_0$ and $\omega_a$. When $\omega_{0} = -1.0$, the range of $\omega_a$ spans $0.94 \leq \omega_a \leq 1.78$, while $\Omega_{\m0}$ ranges from $0.109$ to $0.217$ at the 68\% confidence level.  For the $\omega_a = 0$ model, constraints at the 68\% confidence level are $0.312 \leq \Omega_{\m0} \leq 0.384$, $-1.37 \leq \omega_{0} \leq -1.07$, and $0.610 \leq \textrm{h} \leq 0.798$. 

However, allowing $\omega_a$ to vary yields excessively small values of $\Omega_{\m0}$, making them impractical as viable models. Permitting both $\omega_0$ and $\omega_a$ to vary, the ranges of cosmological parameters at the 68\% confidence level are $-0.78 \leq \omega_0 \leq -0.74$, $1.05 \leq \omega_a \leq 1.55$, and $-0.13 \leq \Omega_{\m0} \leq 0.006$.  Negative values of $\Omega_{\m0}$ arise from statistical analysis, but physically, negative mass is not meaningful in the SMC. Therefore, we exclude consideration of this model. 

Next, we explore meVSL models employing the CPL parameterization of dark energy, where $b \neq 0$.

\subsection{Fixing $\omega_0 = -1$ or $\omega_a = 0$}
\label{subsec:w0m1wa0}

First, fixing $\omega_0 = -1$ and allowing other variables to vary, we obtain the values of $M_0$, $\omega$, $\Omega_{\m0}$, $h$, and $b$ within the 1-$\sigma$ confidence level as ($-19.4216\,,-19.0246$), ($-1.36\,,-1.52$), ($-0.167\,,-0.037$), ($0.68\,,0.816$), and ($0.308\,,0.62$), respectively. Since this model also yields negative $\Omega_{\m0}$, we exclude consideration of this model.

Next, fixing $\omega_a = 0$ and allowing other variables to vary, we conduct a maximum likelihood analysis. In this case, we obtain the values of $M_0$, $\omega$, $\Omega_{\m0}$, $h$, and $b$ within the 1-$\sigma$ confidence level as ($-19.6797\,,-19.4229$), ($-1.26\,,-1.18$), ($0.247\,,0.423$), ($0.606\,,0.682$), and ($-0.154\,,0.198$), respectively. This model exhibits a relatively small value of $M_0$ compared to other models.

\subsection{$\omega_0 = -0.95$ with fixing $\Omega_{\m0}$}
\label{subsec:w0m95}

We perform a maximum likelihood analysis for models with $\omega_0 = -0.95$, while varying $\Omega_{\m0}$ from $0.28$ to $0.32$. Across these models, the best-fit values of h consistently hover around $0.69$, with a 1-$\sigma$ error margin of $0.07$. With increasing $\Omega_{\m0}$, the best-fit value of $\omega_a$ also rises, whereas the best-fit values of $b$ decline.

At a 68\% confidence level, all values of $b$ fall into the negative range for the specified $\Omega_{\m0}$ range, spanning from $-0.344$ to $-0.273$. This suggests that the speed of light decreases monotonically over cosmic time, with its rate of decrease accelerating as $\Omega_{\m0}$ increases.

At $z = 3$, the speed of light can exceed $c_0$ by approximately $10^{13}$\% when $\Omega_{\m0}$ ranges from $0.28$ to $0.32$. The 1-$\sigma$ ranges of $\dot{c}_0/ c_0$ ($10^{-12} \, \text{yr}^{-1}$) are ($-5.05 \,, -4.55$) \,, ($-5.38 \,, -4.88$) \,, ($-5.69 \,, -5.20$) \,, ($-5.99 \,, -5.50$) \,, and ($-6.31 \,, -5.82$), respectively.

\subsection{$\omega_0 = -1.0$ with fixing $\Omega_{\m0}$}
\label{subsec:w0m1}

We conduct an extensive analysis focusing on $\omega_0 = -1.0$ models. Across these models, the best-fit values of h range from $0.689$ to $0.697$ for the specified values of $\Omega_{\m0}$. Meanwhile, the best-fit values of $\omega_a$ and $b$ span $1.73 \leq \omega_a \leq 1.97$ and $-0.293 \leq b \leq -0.215$, respectively.

Similar to the $\omega_0 = -0.95$ models, the best-fit values of $b$ decrease with increasing $\Omega_{\m0}$, with all $b$-values falling into the negative range at a 68\% confidence level. This indicates a monotonically decreasing trend in the speed of light over cosmic time in these models. Notably, compared to the $\omega_0 = -0.95$ model, the $\omega_0 = -1.0$ model exhibits a slightly slower rate of decrease in the speed of light.

At $z = 3$, the speed of light exceeds $c_0$ by approximately $8 \times 10$\% when $\Omega_{\m0}$ ranges from $0.28$ to $0.32$. Furthermore, the 1-$\sigma$ ranges of $\dot{c}_0/ c_0$ ($10^{-12} \, \text{yr}^{-1}$) are ($-4.05 \,, -3.52$) \,, (-$4.45 \,, -3.92$) \,, ($-4.77 \,, -4.24$) \,, ($-5.09 \,, -4.56$) \,, and ($-5.42 \,, -4.89$), respectively.

\subsection{$\omega_0 = -1.05$ with fixing $\Omega_{\m0}$}
\label{subsec:w0m105}

In this subsection, we perform a maximum likelihood analysis for models with $\omega_0 = -1.05$. Across these models, the best-fit values of $h$ consistently hover around $0.69$ for all specified values of $\Omega_{\m0}$. The range of best-fit values for $\omega_a$ spans from $1.57$ to $1.84$, while for $b$, it extends from $-0.240$ to $-0.154$. Similar to previous models, the values of $b$ decrease with increasing $\Omega_{\m0}$, ranging from $-0.169$ to $-0.139$ for $\Omega_{\m0} = 0.28$ and from $-0.255$ to $-0.225$ for $\Omega_{\m0} = 0.32$ within a 1-$\sigma$ error. Thus, the speed of light continues its monotonic decrease over cosmic time in these models. At $z = 3$, $c$ can exceed $c_0$ by approximately $6\%$ ($9\%$) for $\Omega_{\m0} = 0.28$ ($0.32$). Compared to the $\omega_0 = -0.95$ and $-1.0$ models, the rate of decrease in the speed of light is slightly smaller in this model. The 1-$\sigma$ ranges of $\dot{c}_0/ c_0$ ($10^{-12}\, \text{yr}^{-1}$) are ($-2.98\,, -2.45$), ($-3.44\,, -2.82$), ($-3.70\,, -3.17$), ($-4.01\,, -3.49$), and ($-4.50\,, -3.97$), respectively.

\subsection{CPL with fixing $\Omega_{\m0}$}
\label{subsec:cplOm0}

We analyze the Pantheon data without constraining values of $\omega_0$ and $\omega_a$ within the range $0.28 \leq \Omega_{\m0} \leq 0.32$. Across these models, the best-fit values of $h$ remain approximately constant at $0.71$ for all specified values of $\Omega_{\m0}$. The best-fit values of both $\omega_{0}$ and $\omega_a$ range as ($-1.18\,, 0.72$), ($-1.18\,, 0.76$), and ($-1.20\,, 0.71$) for $\Omega_{\m0} = 0.28, 0.30$, and $0.32$, respectively. Of particular interest is the model with $\Omega_{\m0} = 0.30$, where the best-fit value of $b$ is nearly zero, varying within the range $-0.015$ to $0.015$ at a $68$ \% confidence level. This suggests no significant time variation in the speed of light in this model. However, for $\Omega_{\m0} = 0.28$ and $0.32$, the 1-$\sigma$ values of $b$ indicate a clear monotonic decrease (increase) in the speed of light with cosmic time, with ranges of $0.023 \leq b \leq 0.053$ and $-0.039 \leq b \leq -0.009$, respectively.

\begin{figure*}
\centering
\vspace{1cm}
\begin{tabular}{cc}
\includegraphics[width=0.5\linewidth]{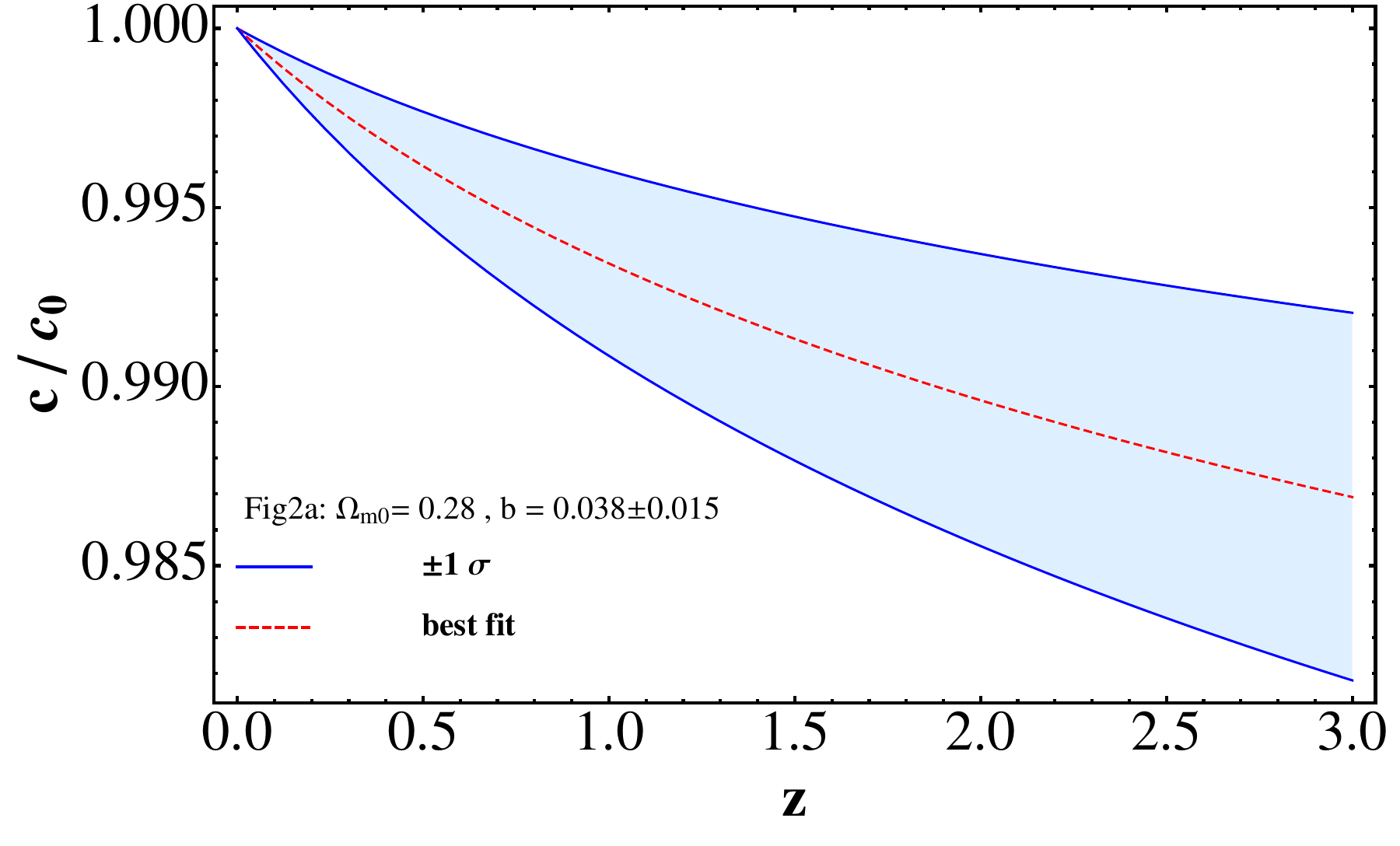} &
\includegraphics[width=0.5\linewidth]{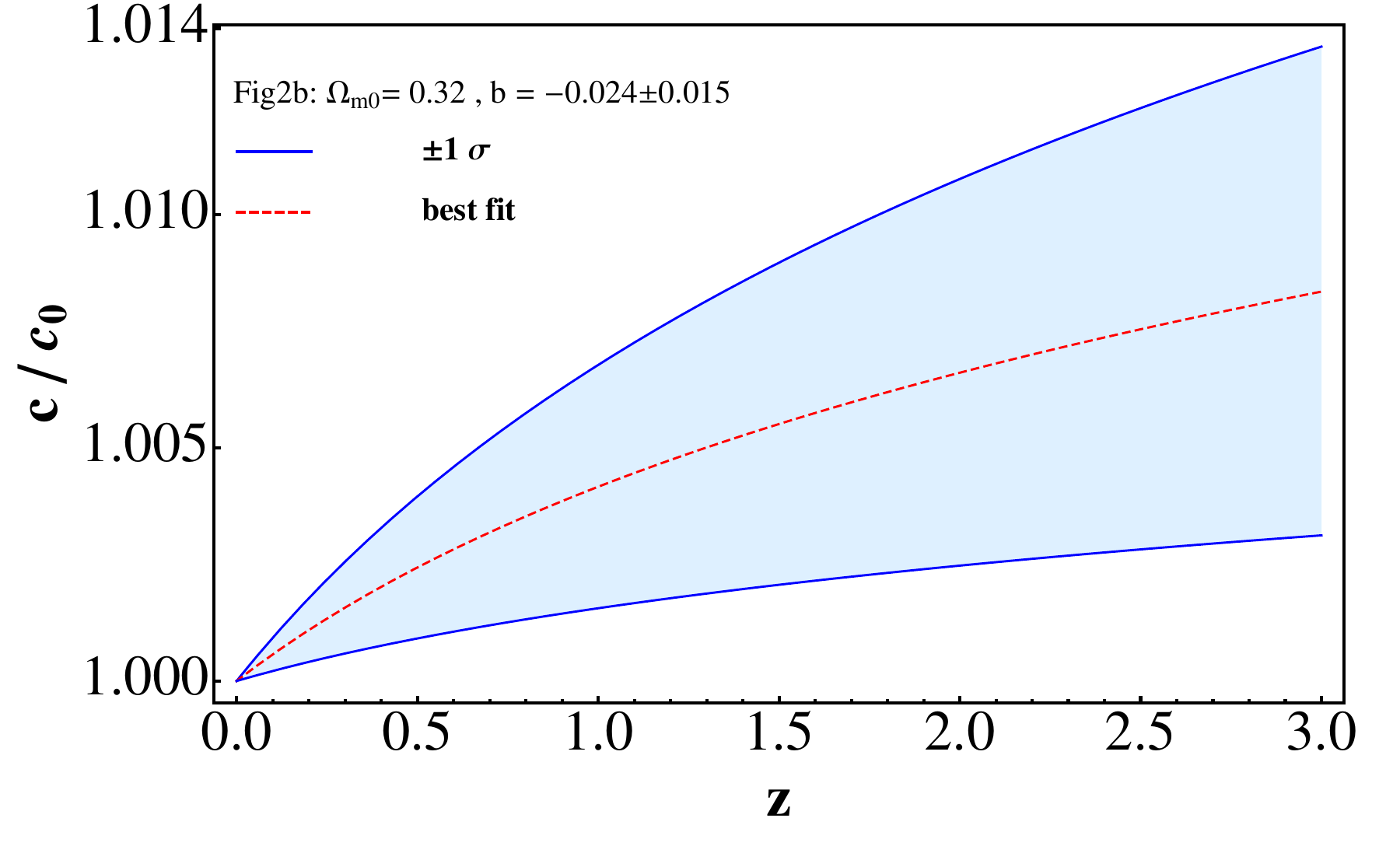}
\end{tabular}
\vspace{-0.5cm}
\caption{The ratios of $c$ to its present value, $c_0$, are plotted as a function of redshift for different values of $\Omega_{\Tm0}$. Panel (a) shows $c/c_0$  for $\Omega_{\m0} = 0.28$, with the dashed line indicating the best-fit value and the solid lines representing the 1-$\sigma$ error range. Similarly, panel (b) displays $c/c_0$ for $\Omega_{m0}  = 0.32$, where the dashed line marks the best-fit value and the solid lines denote the 1-$\sigma$ errors.} \label{fig-2}
\vspace{1cm}
\end{figure*}

These trends are illustrated in Fig.~\ref{fig-2}. In the left panel, the cosmological evolution of $c/c_0$ as a function of $z$ for the $\Omega_{\m0} = 0.28$ model is depicted. The dashed line represents the best-fit value of $b$, while the solid lines indicate the 1-$\sigma$ errors. The monotonically decreasing behavior of $c/c_0$ with increasing redshift is evident due to the positive values of $b$. At $z = 3$, $c$ decreases by approximately $0.8$ ($1.8$) \% within 1-$\sigma$ error. The 1-$\sigma$ range of $\dot{c}_0/c_0$ ($10^{-13} \,\text{yr}^{-1}$) falls between $4.15$ and $9.57$ at a 68 \% confidence level.

On the other hand, the right panel of Fig.~\ref{fig-2} illustrates the model with $\Omega_{\m0} = 0.32$, where both the best-fit and 1-$\sigma$ error values of $b$ are negative. Consequently, $c/c_0$ increases monotonically with redshift. At $z = 3$, $c$ increases by about $0.3$ ($1.4$) \% at a 68 \% confidence level. The 1-$\sigma$ range of $\dot{c}_0/ c_0$ ($10^{-13} \, \text{yr}^{-1}$) spans from $(-7.05 \,, -1.63)$.

%%%%%%%%%%%%%%%%%%%%%%%%%%%%%%%%%%%%%%%
%%%%%%%%%%%%%%%%%%%%%%%%%%%%%%%%%%%%%%% 
  
\subsection{CPL with (without) fixing h}
\label{subsec:cplh}

The analysis is conducted with fixed values of $h$ while leaving other parameters unconstrained. For h$= 0.6736 (0.74)$, the best-fit values and 1-$\sigma$ errors for $\omega_0$ and $\omega_a$ are $-0.99 \pm 0.02 (-0.99 \pm 0.01)$ and $1.31 \pm 0.11 (1.45 \pm 0.07)$, respectively. Notably, in these models, the values of $b$ are positive. However, the resulting values for $\Omega_{\m0}$ fall outside of the viable range, yielding $-0.028 \leq \Omega_{\m0} \leq 0.094 \, (-0.148 \leq \Omega_{\m0} \leq -0.076)$ for $h = 0.6736 \, (0.74)$, which are deemed nonviable. Moreover, without constraining $\Omega_{\m0}$, the matter density contrast is estimated to be $-0.14 \leq \Omega_{\m0} \leq 0.032$, again falling outside of acceptable bounds. These results underscore the importance of appropriately constraining cosmological parameters to ensure the viability of the models.  %\textcolor{blue}{CCC+TL cosmology \cite{Gupta:2024eqo} Scale Invariant Vacuum (SIV) paradigm \cite{Gueorguiev:2022wit}}

\section{$\dot{c}$ and $\dot{G}$ }
\label{sec:dotcdotG}

Expanding on the insights gained from previous subsections~\ref{sec:wCDM} and \ref{sec:CPL}, we embark on an exploration of viable meVSL models across diverse dark energy scenarios, aiming to extract valuable constraints on both cosmological and model parameters. This analysis allows us to derive estimates for the temporal evolution of the speed of light, a fundamental aspect within the meVSL framework. Notably, in the meVSL paradigm, the speed of light undergoes cosmological evolution, mirroring the behavior of the gravitational constant, which is characterized by $G = G_0 (1+z)^{-b}$. By leveraging the constraints obtained on $b$-values, we can ascertain bounds on the present value of the relative temporal variation of the gravitational constant, $\dot{G}_0/G_0$.

To place these findings within the broader context of observational constraints, we compare our results with existing bounds on $\dot{G}_0/G_0$ from various sources, as summarized in Table~\ref{tab:dotGRef}. Notably, the analysis of lunar laser ranging (LLR) data stands out for its stringency, yielding the most stringent bounds on $\dot{G}_0/ G_0$. In contrast, the orbital period rate of pulsars offers the widest bounds, estimated at $2.3 \times 10^{-11} \, \text{yr}^{-1}$. Taken together, these observations suggest that $\dot{G}_0/G_0$ is tentatively within the order of $10^{-12} \, \text{yr}^{-1}$. Such insights not only deepen our understanding of cosmological dynamics but also pave the way for future investigations into the fundamental nature of physical constants.

 \begin{table}[h!]
 	\begin{center}
		\caption{Table provides the latest 1-$\sigma$ observational constraints on the present rate of change of the gravitational constant, $\dot{G}_0/G_0$. Here, "WD" refers to white dwarf observations, "BBN" signifies Big Bang nucleosynthesis, "LLR" denotes lunar laser ranging data, and "GWs" represents gravitational waves.}
		\label{tab:dotGRef}
		\begin{tabular}{|c|c|c|} 
			\hline
			obs  & $\dot{G}_0/G_0 \,(10^{-12} \, \text{yr}^{-1})$ & Ref  \\ \hline
			pulsars  & 23 & \cite{Verbiest:2008gy} \\ \hline
			\multirow{2}{*}{WD} cooling  &-1.8  &  \cite{GarciaBerro:2011wc} \\ 
			\quad \quad \,\, pulsation  & -130 & \cite{Corsico:2013ida} \\ \hline
			\multirow{2}{*}{BBN} & $-0.3 \sim 0.4$ & \cite{Bambi:2005fi} \\
				& $-3.6 \sim 4.5$ & \cite{Alvey:2019ctk} \\ \hline
			\multirow{2}{*}{LLR} & $-0.5 \sim 0.9$ & \cite{Hofmann:2010} \\
				& $-0.005 \sim 0.147$ & \cite{Hofmann:2018} \\ \hline
			\multirow{2}{*}{SNe Ia} & $-30 \sim 73$ & \cite{Mould:2014iga} \\
				& $3$ & \cite{Zhao:2018gwk} \\ \hline
			\multirow{2}{*}{GWs} LIGO& $70$ & \cite{Lagos:2019kds} \\
			\quad \,\,\,\,\,\, LISA & $0.7$ & \cite{Belgacem:2019pkk}\\ \hline
\end{tabular}
	\end{center}
 \end{table}

Table~\ref{tab:dotcG} displays the outcomes regarding the temporal variations of both the speed of light and the gravitational constant within meVSL models, considering various dark energy scenarios. We define $\Delta c(z=3)$ as the percentage deviation between the speed of light's value at redshift $z = 3$ and its current value, expressed as $\Delta c(z=3) \equiv \left( c(z=3) - c_0 \right)/c_0 \times 100 \, (\%)$ within a 1-$\sigma$ uncertainty range. Similarly, $\Delta G(z=3)$ indicates the percentage deviation between the gravitational constant's values at $z = 3$ and $z = 0$. The ratio of the temporal variation of the speed of light to its current value is denoted by $\dot{c}_0/c_0$, while $\dot{G}_0/G_0$ represents the current ratio of the gravitational constant's temporal variation to its value.

In the analysis, positive values of the best-fit parameter $b$ and its 68\% confidence level values are only observed for the CPL dark energy model when $\Omega_{\m0} = 0.28$. Consequently, both $\dot{c}_0/c_0$ and $\dot{G}_0/G_0$ exhibit positivity in this specific model. Conversely, all other viable models derived from the Pantheon data yield negative $b$ values, resulting in negative values for both $\dot{c}_0/c_0$ and $\dot{G}_0/G_0$ in these scenarios.

The values of $\dot{c}_0/c_0$ are around $10^{-13} \, \text{yr}^{-1}$, whereas $\dot{G}_0/G_0$ values are approximately $10^{-12} \, \text{yr}^{-1}$ for $\omega$CDM models and CPL models with varying $\omega_0$ and $\omega_a$, as detailed in Table~\ref{tab:dotcG}. However, for CPL models with fixed $\omega_0$, the $\dot{c}_0/c_0$ values are roughly $10^{-12} \, \text{yr}^{-1}$, while $\dot{G}_0/G_0$ values tend to be around $10^{-11} \, \text{yr}^{-1}$.

Despite having three different constraints on $\dot{G}_0/G_0$ derived from SNe Ia data, including this work, there are notable discrepancies among the results \cite{Mould:2014iga, Zhao:2018gwk}. These discrepancies stem from the varying methodologies and datasets used. For instance, one approach employs the standard candle method to define the redshift-distance relation, establishing a broad upper limit on $| \dot{G} / G|$ and exploring various parameterizations and impacts \cite{Mould:2014iga}. In contrast, another method focuses on the intrinsic properties of Type Ia supernovae (SNIa) and their dependency on the Chandrasekhar mass, $M_{\textrm{Ch}} \propto G^{-3/2}$, to track the variation of $G$ across different redshifts \cite{Zhao:2018gwk}. Expanding on the insights gained from previous subsections~\ref{sec:wCDM} and 3, we embark on an exploration of viable meVSL models across diverse dark energy scenarios, aiming to extract valuable constraints on both cosmological and model parameters. This analysis allows us to derive estimates for the temporal evolution of the speed of light, a fundamental aspect within the meVSL framework. Notably, in the meVSL paradigm, the speed of light undergoes cosmological evolution, mirroring the behavior of the gravitational constant, which is characterized by $G = G_0 (1+z)^{-b}$. By leveraging the constraints obtained on $b$-values, we can ascertain bounds on the present value of the relative temporal variation of the gravitational constant, $\dot{G}_0/G_0$.

 \begin{table}[h!]
		\caption{Time variations of the speed of light and that of the gravitational constant at a 68 \% confidence level for viable models. We denote $A \equiv 10^{-13} \,\text{yr}^{-1}$}
		\label{tab:dotcG}

\begin{adjustwidth*}{}{}
\footnotesize
%\centering %% If there is a figure in wide page, please release command \centering
 	\begin{center}
		\begin{tabular}{|c|c|c|c|c|c|c|c|c|} 
			\hline
			 $\omega_0$& $\omega_a$ & $\Omega_{\m0}$ & $\Delta c (z=3) \, (\%)$ & $\dot{c}_0/c_0 \, [A]$ & $\Delta G(z=3) \, (\%)$  & $\dot{G}_0/G_0 \, [10A]$ & $M_0$ & $b$ \\ \hline
			 $\multirow{3}{*}{-1}$ & $\multirow{3}{*}{0}$ & $0.30$ & $0.2 \sim 1.7$ & $-8.76 \sim -0.89$ & $0.7 \sim 7.0$  & $-3.51 \sim -0.36$ & $-19.3561 \pm 0.2899$ & $-0.027 \pm 0.022$ \\ 
			 $$ & $$ & $0.31$ & $0.8 \sim 2.3$ & $-1.18 \sim -0.39$ & $3.1 \sim 9.6$  & $-4.72 \sim -1.57$  & $-19.3563 \pm 0.2899$ & $-0.044 \pm 0.022$ \\  
			 $$ & $$ & $0.32$ & $1.4 \sim 2.9$ & $-1.48 \sim -0.70$ & $5.6 \sim 12.2$  & $-5.94 \sim -2.79$ & $-19.3566 \pm 0.2898$ & $-0.061 \pm 0.022$ \\ \hline
			 $\multirow{3}{*}{-0.95}$ & $1.85 \pm 0.22$ & $0.28$ & $9.4 \sim 10.5$ & $-51.3 \sim -46.3$ & $43.2 \sim 48.9$  & $-20.5 \sim -18.5$  & $-19.4115 \pm 0.2313$ & $-0.273 \pm 0.014$ \\ 
			 $$ & $1.95 \pm 0.23$ & $0.30$ & $10.8 \sim 11.8$ & $-57.8 \sim -52.8$ & $50.5 \sim 56.5$  & $-23.1 \sim -21.1$ & $-19.4110 \pm 0.2328$ & $-0.309 \pm 0.014$ \\ 
			 $$ & $2.07 \pm 0.24$ & $0.32$ & $12.1 \sim 13.2$ & $-64.0 \sim -59.0$ & $58.0 \sim 64.3$  & $-25.6 \sim -23.6$ & $-19.4119 \pm 0.2342$ & $-0.344 \pm 0.014$ \\ \hline
			 $\multirow{3}{*}{-1.0}$ & $1.73 \pm 0.23$ & $0.28$ & $7.2 \sim 8.3$ & $-41.1 \sim -35.8$ & $32.0 \sim 37.6$  & $-16.5 \sim -14.3$  & $-19.4108 \pm 0.2322$ & $-0.215 \pm 0.015$  \\ 
			 $$ & $1.85 \pm 0.24$ & $0.30$ & $8.7 \sim 9.8$ & $-48.3 \sim -42.9$ & $39.5 \sim 45.4$  & $-19.3 \sim -17.2$ & $-19.4036 \pm 0.2344$ & $-0.255 \pm 0.015$ \\ 
			 $$ & $1.97 \pm 0.25$ & $0.32$ & $10.1 \sim 11.3$ & $-55.1 \sim -49.7$ & $47.0 \sim 53.2$  & $-22.0 \sim -19.9$  & $-19.4107 \pm 0.2353$ & $-0.293 \pm 0.015$ \\ \hline
			 $\multirow{3}{*}{-1.05}$ & $1.57 \pm 0.24$ & $0.28$ & $4.9 \sim 6.0$ & $-30.2 \sim -24.9$ & $21.3 \sim 26.4$  & $-12.1 \sim -9.9$ & $-19.4038 \pm 0.2339$& $-0.154 \pm 0.015$ \\ 
			 $$ & $1.68 \pm 0.25$ & $0.30$ & $6.4 \sim 7.5$ & $-37.6 \sim -32.2$ & $28.3 \sim 33.8$  & $-15.0 \sim -12.9$ &$-19.4079 \pm 0.2351$ & $-0.195 \pm 0.015$ \\ 
			 $$ & $1.84 \pm 0.26$ & $0.32$ & $8.1 \sim 9.2$ & $-45.6 \sim -40.2$ & $36.6 \sim 42.4$  & $-18.2 \sim -16.1$ &$-19.4082 \pm 0.2367$ & $-0.240 \pm 0.015$  \\ \hline
			 $-1.18 \pm 0.04$ & $0.72 \pm 0.23$ & $0.28$ & $0.8 \sim 1.8$ & $4.11 \sim 9..48$ & $3.1 \sim 7.1$  & $1.65 \sim 3.79$ & $-19.3499\pm0.1176$ & $0.038\pm0.015$ \\ 
			 %$-1.18 \pm 0.04$ & $0.76 \pm 0.25$ & $0.30$ & $6.4 \sim 7.5$ & $-37.6 \sim -32.2$ & $28.3 \sim 33.8$  & $-15.0 \sim -12.9$  \\ 
			 $-1.20 \pm 0.04$ & $0.71 \pm 0.26$ & $0.32$ & $0.3 \sim 1.4$ & $-6.98 \sim -1.61$ & $1.3 \sim 5.6$  & $-2.79 \sim -0.64$ &$-19.3478 \pm 0.1225$ & $-0.024 \pm 0.015$ \\ \hline
\end{tabular}
	\end{center}
\end{adjustwidth*}
 \end{table}

%%%%%%%%%%%%%%%%%%%%%%%%%%%%%%%%%%%%%%%%%%
\section{Discussion}
\label{sec:conclusion}

The Pantheon$+$ data provides constraints on cosmological and model parameters with a statistical precision of about 10\%. Leveraging this dataset, we perform a maximum likelihood analysis to constrain dark energy models within the framework of the modified varying speed of light (meVSL) model. This analysis allows us to identify several viable $\omega$CDM and CPL dark energy models and derive constraints on the parameter $b$, which governs the evolution of physical constants in the Universe. 

The constraints obtained from our analysis indicate that the relative temporal variations of the speed of light and the gravitational constant lie within the ranges $-64.0 \leq \dot{c}_0/c_0 \, (10^{-13} \,\text{yr}^{-1}) \leq -0.39$ and $-25.6 \leq \dot{G}_0/G_0 \, (10^{-12} \, \text{yr}^{-1}) \leq -0.36$, respectively, for most viable models. These findings suggest that, according to the current Pantheon data, both the speed of light and the gravitational constant were greater in the past and have decreased monotonically with redshift, $z$.

Among the CPL models we considered, some do not require a dark matter energy density. We understand this phenomenon as a result of the degeneracy between $\omega_a$ and $\Omega_{\m0}$. We focused our viable model consideration on meVSL models where the cosmological parameters closely match those of the Standard Cosmological Model (SCM). However, models with $\Omega_{\m0}$ approaching zero could become an interesting topic of study, requiring a deeper understanding and further research. For the purposes of this paper, we defer to the references cited within the main text to substantiate these points.

While additional cosmological observations, such as those from the Cosmic Microwave Background (CMB) and Baryon Acoustic Oscillations (BAO), could provide further constraints on cosmological and model parameters, integrating these datasets would necessitate a reanalysis within the theoretical framework of the meVSL model. This comprehensive task falls outside the scope of the present manuscript and is deferred to future investigations. Future studies incorporating CMB and BAO data could significantly refine the constraints on the parameters governing the evolution of physical constants, thereby enhancing our understanding of the meVSL model and its implications for cosmology.

%%%%%%%%%%%%%%%%%%%%%%%%%%%%%%%%%%%%%%%%%%%
\section*{Acknowledgments}
%%%%%%%%%%%%%%%%%%%%%%%%%%%%%%%%%%%%%%%%%%%
This research was funded by the National Research Foundation of Korea (NRF), funded both by the Ministry of Science, ICT, and~Future Planning (Grant No. NRF-2019R1A6A1A10073079) and by the Ministry of Education (Grant No. NRF-RS202300243411). S. L. thanks the editor for the invitation to publish this manuscript in the Universe and for their helpful and constructive comments.

%\institutionalreview{}

%\informedconsent{}

%\dataavailability{No new data were created and used in this research.} 

%\conflictsofinterest{The authors declare no conflicts of interest.} 

%\appendixtitles{no} % Leave argument "no" if all appendix headings stay EMPTY (then no dot is printed after "Appendix A"). If the appendix sections contain a heading then change the argument to "yes".
%\appendixstart

\end{document}